\documentclass[aps,onecolumn,preprint,superscriptaddress,nofootinbib,floats]{revtex4}
\usepackage{amsmath,amssymb,color,mathrsfs, graphicx,verbatim,epsfig, bbm, wasysym, natbib}

\def\me{m_{\tilde{e}}}
\def\ml{m_{\tilde{l}}}
\def\mB{m_{\tilde{B}}}

\def\Ye{Y_{\tilde{e}}}
\def\Yl{Y_{\tilde{l}}}

\def\gaugino{\tilde{\chi}}
\def\stau{\tilde{\tau}}

\def\beq{\begin{equation}}
\def\eeq{\end{equation}}
\def\beqarray{\begin{eqnarray}}
\def\eeqarray{\end{eqnarray}}

\def\bit{\begin{itemize}}
\def\eit{\end{itemize}}

\def\bino{\tilde{B}}
\def\wino{\tilde{W}}
\def\gluino{\tilde{g}}
\def\slep{\tilde{l}}
\def\snu{\tilde{\nu}}
\def\sel{{\tilde{e}}_R}
\def\squark{\tilde{q}}

\def\higgsino{\tilde{H}}

\def\met{$\displaystyle{\not}E_T$}

\def\simgt{\mathrel{\lower2.5pt\vbox{\lineskip=0pt\baselineskip=0pt
           \hbox{$>$}\hbox{$\sim$}}}}
\def\simlt{\mathrel{\lower2.5pt\vbox{\lineskip=0pt\baselineskip=0pt
           \hbox{$<$}\hbox{$\sim$}}}}

\begin{document}
\begin{flushright}UMD-PP-10-007
\end{flushright}
\vspace{0.2cm}

\title{Leptophilic Signals of a Sneutrino (N)LSP and \\
Flavor Biases from Flavor-Blind SUSY}

\author{Andrey Katz}
\email{andrey(at)umd.edu}
\affiliation{Department of Physics, University of Maryland,\\
College Park, MD 20742}

\author{Brock Tweedie}
\email{brock(at)pha.jhu.edu}
\affiliation{Department of Physics and Astronomy, Johns Hopkins University,\\
Baltimore, MD 21218}

\date{\today}

\begin{abstract} 
\noindent Although the sneutrino is a viable NLSP candidate with 
gravitino LSP, spectra of this type occupy a part of SUSY parameter 
space in which collider signatures are poorly studied.  In this 
paper we will extend previous work on this topic
to include sneutrino NLSP spectra with non-minimal phenomenology. 
Generally, these spectra exhibit very leptophilic behavior, 
which can be easily observed at the LHC.  We show that a 
variety of such spectra
can be analyzed with similar techniques, leading in each case to very suggestive
evidence for complicated decay chains that end in sneutrinos. 
Amongst the variations considered, we find a simple class of
spectra that produce signals with strong electron-muon asymmetries.
These signals could naively be interpreted as evidence for lepton
flavor violation, but can occur even with flavor-blind SUSY.
\end{abstract}

\maketitle

\section{Introduction}

The phenomenology of the Minimal Supersymmetric Standard Model (MSSM) has been an 
evolving topic of investigation for many years.
While traditional supersymmetry (SUSY) mediation scenarios continue to remain relevant, 
supersymmetric signals have failed to manifest themselves over the course of several generations
of colliders and precision low-energy experiments.  As the constraints on the soft
parameters have progressively tightened, SUSY model-building has proliferated, realizing 
unexplored and sometimes surprising regions of the MSSM parameter space. 
Given the vast number of possibilities, it is clear that we should be prepared for anything
as we enter the LHC era, and that general classification of novel signals is a useful 
endeavor, independent of any specific UV motivation.

With this in mind, we have initiated an inquiry into the phenomenology of a broad, 
relatively unstudied region of the MSSM, in which the sneutrino acts like the lightest
supersymmetric particle (LSP) at 
the LHC~\cite{Katz:2009qx} in place of more standard options such as a neutralino.  
Such spectra have been largely neglected in part because the sneutrino is excluded as a stable
thermal relic of the big bang~\cite{Falk:1994es, Arina:2007tm}, and also because it tends to be heavier 
than other superparticles in minimal mediation scenarios.  However, spectra with sneutrino NLSP and 
gravitino LSP are quite cosmologically safe within a standard thermal 
history~\cite{Feng:2003xh,Buchmuller:2006nx,Kanzaki:2006hm}.
We have also seen that
obtaining a sneutrino NLSP is straightforward both in low-scale mediation frameworks such as General
Gauge Mediation~\cite{Meade:2008wd,Carpenter:2008he,Rajaraman:2009ga},
as well as in high-scale 
mediation~\cite{Kaplan:2000av,Buchmuller:2005ma,Ellis:2002iu,Ellis:2008as,Kadota:2009vq}.

Since it decays invisibly, the sneutrino NLSP is indistinguishable from an LSP within a collider, 
leading us to informally dub it an ``(N)LSP.''  Indeed, it would be unclear from collider data
alone whether such a particle is the NLSP versus the true LSP within a nonstandard cosmology, or 
whether it decays via $R$-parity-violating interactions after exiting the detector.  
In~\cite{Katz:2009qx}, we found that this class of scenarios can lead to very distinctive signatures, 
even within the simplest spectra.  Our discussion 
there was limited to cases with flavor-degeneracy, small $A$-terms, squarks/gluinos 
heavier than at least one neutralino with sizable gaugino component, 
and right-handed (RH) sleptons playing almost no role in the cascades.

Given these assumptions, every cascade typically has a roughly 30\% probability of producing an 
electron or muon, independent of which electroweak gauginos participate.  This is due to the fact 
that the NLSP sneutrino and its $SU(2)$ partner, the charged left-handed (LH) slepton, are 
generally quasi-degenerate, and will be produced approximately democratically in gaugino decay.  
In addition, the decays of the charged LH sleptons can also produce 
leptons via $W^*$ emission.  In chains where the LH slepton is produced in a neutralino decay and 
subsequently decays leptonically, the two leptons produced will be correlated 
in charge but uncorrelated in flavor.  Together, these features lead to three distinguishing signals:
\begin{enumerate}
\item high rates for multilepton events up to trilepton and even 4-lepton,
\item a flavor-blind excess of 
opposite-sign (OS) dileptons over same-sign (SS) dileptons with a distinctive invariant mass distribution,
\item the same OS dilepton kinematic feature contained within the trilepton sample.
\end{enumerate}

Here, we seek to explore some variations on this basic scenario.  First,
we will relax the assumption that the RH sleptons are effectively decoupled from phenomenology.
This means that we will consider spectra where $m(\sel) < m(\bino)$, and in which
production of the mostly-Bino neutralino in squark/gluino cascades is not too small. We call these 
``RH-active'' spectra.  Decay chains with
RH sleptons will contain additional electron and muon emissions, leading to a richer structure
of multileptonic signals.  Besides overall higher lepton multiplicities---with significant three-, 
four-, and five-lepton signals---there will be several overlapping kinematic distributions in
most channels.  We note that these spectra share much 
in spirit with the ``leptogenic'' SUSY scenario
of~\cite{DeSimone:2008gm,DeSimone:2009ws}, though with the ordering of RH versus LH sleptons 
reversed, and without CHAMPs (CHArged Massive Particles).  In particular, every event will end with 
two sneutrinos instead
of two quasi-stable RH sleptons, introducing different decay topologies and limiting all kinematic 
reconstructions.  

Interestingly, RH-active spectra can also lead to highly flavor-biased signals within the first two 
generations, in spite of the fact that we continue to work exclusively with flavor-blind SUSY mediation
with small $A$-terms.
In some cases, left-right mixing effects can become important in the decays of not only the 
(mostly-)RH stau, but also the RH smuon.  This can radically alter the decay chains with RH smuons with 
respect to RH selectrons, despite the fact that these sleptons will be nearly mass degenerate.  For
example, a RH smuon can mix into a LH smuon, and decay directly to the NLSP sneutrino by emitting a 
real or virtual $W$.  When such mixing-induced smuon decays dominate, they will lead to
${\mathcal O}(1)$ flavor non-universality in the multilepton signals.  At first sight, the
observation of such signals at the LHC could be interpreted as highly flavor non-universal 
slepton soft terms, similar to~\cite{Kribs:2007ac,Feng:2007ke,Nomura:2007ap,Kumar:2009sf}.  
However, in our spectra, the non-universality originates entirely with the
(supersymmetrized) Standard Model Yukawa couplings at tree level.

The last variation which we consider here is the class of spectra where the LH stau doublet becomes split
off from the first two generations, so that $m(\snu_\tau) < m(\tilde\tau_L) < m(\snu_{e,\mu})$.
This can happen due to Yukawa-dependent running effects in high-scale models.  In the specific
case that we study, all of the slepton doublets 
become light due to $D$-term corrections induced by a large down-type Higgs soft
mass.  The large soft mass further drives the flavor splitting.  
This possibility was pointed out 
in~\cite{Kaplan:2000av,Ellis:2002iu,Buchmuller:2005ma,Ellis:2008as}, and some of the
details of its LHC phenomenology were further
discussed in~\cite{Covi:2007xj,Medina:2009ey,Santoso:2009qa}.\footnote{Parenthetically, we note that such 
scenarios may be subject to constraints from $\mu \to e \gamma$, but this 
depends in detail on the structure of the right-handed neutrino sector.}  Spectra of this 
type are referred to as 
``NUHM'' spectra, for
 non-universal Higgs mass boundary conditions.  The fact that the NLSP (tau sneutrino) and 
NNLSP (mostly-LH stau) both
carry tau-number suggests that the products
of decay chains will be enriched with taus, and indeed this has been the favored collider
signature discussed in the literature~\cite{Covi:2007xj,Medina:2009ey,Santoso:2009qa}.  
Here, we will
see that the analysis techniques which we develop for more flavor-degenerate spectra
can also be applied in this case, independent of the efficiency for identifying taus.

In summary, then, we will be extending our analysis of the simplest sneutrino NLSP 
spectra~\cite{Katz:2009qx} to incorporate
the following possibilities:  a non-negligible role for RH sleptons in SUSY 
decay chains, either with or without flavor-dependent decays of the RH 
sleptons due to left-right mixing, and a significantly lighter stau doublet due 
to running effects from a large down-type Higgs soft mass.  In all of 
these cases, every SUSY decay chain has several new opportunities to produce electrons and muons,
leading to high rates for multilepton signals up to quite high multiplicity.  We continue to concentrate on
distributions within the dileptonic and trileptonic channels, as these have manageable
combinatoric ambiguities and good statistics.  We will see that these spectra can have
significant excess of opposite-sign same-flavor (OSSF) leptons, which typify more standard
spectra with a neutralino LSP.  However, this excess will coexist with the characteristic flavor-uncorrelated OS
signal of LH slepton production and decay, leading to independent excesses in OSSF and 
opposite-sign opposite-flavor (OSOF), mismatched in normalization, shape, and in some cases
electron versus muon composition.  Trilepton will display an additional excess originating
from chains which proceed sequentially through RH sleptons and charged LH sleptons.  Again,
sign and flavor information will serve as useful indicators.  Together, the coexistence
of all of these signals will be quite suggestive of spectra with a sneutrino NLSP, beyond
the simplest cases discussed in~\cite{Katz:2009qx}.

The paper is organized as follows.  In section~\ref{decays}, we discuss the decays
of LH and RH sleptons.  In section~\ref{Signals}, we  show how the presence 
of RH sleptons or a light stau doublet
modifies the multilepton signals characteristic of simpler sneutrino NLSP 
scenarios, possibly in flavor non-universal ways.    
We analyze several representative examples  in simulation in section~\ref{Simulations}.  
Section~\ref{conclusions} contains conclusions 
and discussion. Some technical details are relegated to the appendix.

\section{Decays of the Sleptons}\label{decays}

Our focus in this paper will be the multileptonic signals of sneutrino NLSP spectra beyond
the simplest models.
The leptons are dominantly produced in one of two ways:  in the decays
of gauginos to sleptons, and in the subsequent decay of the sleptons.  Up to chargino and
neutralino mixing effects, which we assume to be modest,\footnote{As we will later 
see from simulation (section~\ref{Simulations}), this assumption is not strictly necessary, 
and the behavior of the 
leptonic channels is qualitatively unchanged even in spectra with highly mixed neutralinos. 
However, we will continue to make this assumption for the purpose of simplifying the discussion.} 
the former production mechanism
is quite simple.  However, slepton decays can be multifaceted in these spectra, and 
here we will dedicate some discussion to these decays.

First we will briefly review the main decays of interest for the LH sleptons.  We then 
move on to discuss the simplest, flavor universal decays of 
RH sleptons.  Finally, we discuss RH slepton decay modes induced by left-right mixing.  These
modes will often dominate RH stau decays, but can also dominate RH smuon decays, leading to
very striking flavor non-universal signals in the first two generations.

\subsection{Left-Handed Sleptons}\label{lhdecays}

The decays of the LH sleptons, $SU(2)$ partners of the NLSP sneutrinos, were seen to
be a crucial ingredient in our original study~\cite{Katz:2009qx}.  The situation is
unchanged in the RH-active spectra which we consider here, but we present
a brief review.  We then proceed to discuss potentially relevant variations.

\begin{figure}[t]
\centering
\includegraphics[width=6.3in]{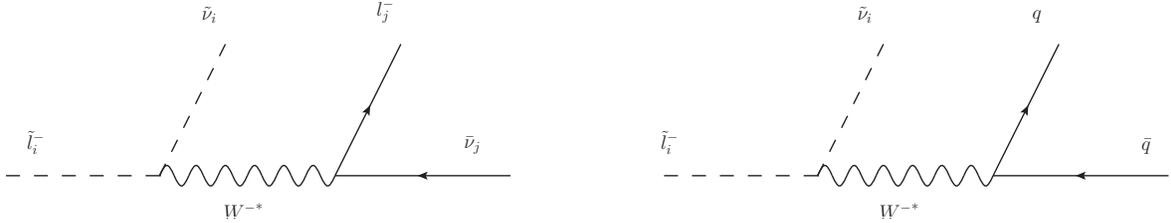}
\caption{Possible decay modes of the LH slepton through $W^*$ emission.}
\label{fig:1}
\end{figure}

Recall that the splitting between the LH sleptons and the sneutrinos is given mostly by
$D$-term interactions with the Higgs VEVs:
\beq
m_{\slep}-m_{\snu} \approx \frac{m_W^2 (-\cos 2\beta)}{m_{\slep}+m_{\snu}}~.
\eeq  
One can clearly see that the splitting cannot exceed the mass of the $W$, and that any decay of the LH 
slepton is necessarily three-body.  For example, for a doublet mass of 200~GeV, and $\tan\beta \simgt 3$, 
the splitting is about 16~GeV.  Typically, the dominant diagram is the 
familiar electroweak decay via $W^*$ emission, as in Fig.~\ref{fig:1}.  Branching fractions 
are as usual.  67\% of decays produce jets, which are in this case relatively soft and low-multiplicity
due to the smallness of the available energy.  These decays will likely be quite difficult to isolate
at the LHC.  11\% of decays produce a tau, also a challenging signal, and sometimes indistinguishable
from a prompt electron or muon production.  The remaining 22\% of decays result in a relatively clean 
electron or muon.

This lepton is completely uncorrelated in flavor with its parent slepton, and with any 
charged lepton produced along with the slepton.  Therefore, decay chains with an intermediate
LH slepton can feature a pair of opposite-sign leptons with no relative flavor structure.  As
discussed in~\cite{Katz:2009qx}, this leads to equal excesses of OSSF and OSOF lepton pairs,
which can be seen in dilepton as well as trilepton events at the LHC.  We argued that the
observation of these unconventional signals would serve as strong evidence for a spectrum 
with sneutrino NLSP where RH sleptons are largely bypassed.

\begin{figure}[t]
 \centering 
\includegraphics[width=6.4in]{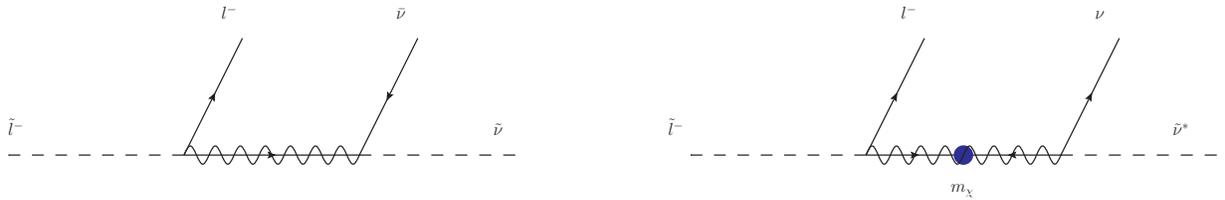}
\caption{Additional neutralino diagrams contributing to LH slepton decays. The diagram on the left 
interferes with the $W^*$ leptonic mode.  The diagram on the right proceeds via a chirality-flip on 
the neutralino propagator.}
\label{fig:2}
\end{figure}

More generally, the $W^*$ diagrams 
are accompanied by other diagrams involving virtual charginos, neutralinos, and, in the case of 
staus, heavy charged Higgses.  Fig.~\ref{fig:2} shows the neutralino diagrams, which consist of
two different, mutually non-interfering cases depending on chirality flow (i.e., whether the $\slep_L$ 
decays into $\snu$ or $\snu^*$).  These can both usually be 
neglected to first approximation, as long as the mass-squared difference
between the neutralinos and the LH sleptons is significantly larger than $m_W^2$.  Equivalently, 
the LH doublet should be split off from the neutralinos by an amount ${\mathcal O}(1)$ 
times larger than its own internal $D$-term mass 
splitting.  In this case, the largest correction to the $W^*$ decays naively comes from the 
chirality-flipping process 
on the right side of Fig.~\ref{fig:2}, since its rate is enhanced by $m_{\gaugino}^2/m_{\slep_L}^2$ 
with respect to the chirality preserving process.  Both of these would introduce an additional population
of OSSF dileptons, since the flavors of the slepton and final lepton are now correlated.   However, 
the chirality-preserving diagrams interfere with the $W^*$ diagrams in the case of same-flavor decays,
leading to larger effects than the chirality-flipping diagrams.  
The interference term can either enhance or deplete the OSSF signals, 
depending on whether the neutralino is mostly-Bino or mostly-Wino, respectively.  We will not
study cases where this effect is large enough to be easily observed, but we note that 
their dilepton signals can be somewhat similar to those of
the NUHM spectra which we do study, if the closest neutralino is Bino-like.  The case of 
interference with a mostly-Wino neutralino
would, however, be quite distinctive, as it would feature an OSSF {\it deficit} from the destructive
interference in same-flavor decays.\footnote{For a mostly-Wino, the relative correction to same-flavor
decays from interference goes like $-m_W^2/(m_{\wino}^2-m_{\slep}^2)$.  (For the Bino, we would make a sign
flip and multiply by $g_1^2/g_2^2$.)  For example, for a 200 GeV slepton and 260 GeV Wino, the correction
is roughly $-$20\%.}

In spectra with a light stau doublet, the effects of these additional diagrams become important for 
a different reason: for a LH slepton of the first two generations, 3-body decays into 
$\snu_\tau \nu_\tau$, as well as  $\stau\tau$, can
have much larger available phase space.  Chirality-flipping and chirality-preserving processes  
each contribute significantly, leading to an observable OSSF excess beyond the OS excess from the 
usual $W^*$ decays.  We will study spectra of this type in more detail below.

We note that we 
will explicitly be assuming that the gap between the tau-sneutrino and the first two
generations of LH sleptons only
contains the LH stau and the electron- and muon-sneutrinos.  In other words, there are no
chargino or neutralino states in between which could cause additional cascades.  If this were
the case, then we would likely lose any indication of the presence of the heavier sneutrinos
in the decay chains, as their partner sleptons would more likely bypass them to undergo 
2-body decays.\footnote{There 
are also several similar variations on the LH slepton decays in the flavor-degenerate
case, which we do not pursue in
detail because they are rather special.  However, we mention the simplest of these for completeness.  
A trivial possibility is the presence of a neutralino sitting in the small mass gap between the LH
sleptons and their sneutrino partners.  This would typically lead to nearly 100\% same-flavor decays
into lepton and neutralino, bypassing the sneutrino.
This ``accidental'' scenario would be quite difficult to discriminate from more traditional 
neutralino LSP spectra.
Another possibility is a low SUSY-breaking scale, so that the LH slepton might directly decay to lepton
and gravitino, again bypassing the sneutrino.  For example, with a 200 GeV (300 GeV)
sneutrino in the large $\tan\beta$ limit, these decays would be comparable to the leptonic $W^*$ decays
for $\sqrt{F} \sim 10$ TeV $(30$ TeV)~\cite{Katz:2009qx}.
While these values are relatively small, it is in principle possible for the gravitino decays to have
observable effects even for ${\mathcal O}(1)$ larger scales.  The dilepton signals would again feature
an additional OSSF excess from these new decay channels.}

Of course, we may also consider taking any of the above variations in LH slepton decay patterns, and
embedding them into a spectrum where RH sleptons are also important.  Such scenarios could have
independent OSSF contributions from LH and RH sleptons, but would otherwise largely appear identical
to the RH slepton cases which we study below.

\subsection{Right-Handed Sleptons: Flavor Universal Decays}\label{rhdecays}

The RH sleptons must ultimately decay down to the LH slepton doublet.  If we
neglect Yukawas, the only available options are real or virtual neutralino 
emissions, coupling through the Bino component.  
While the production and decay of a RH slepton is guaranteed to produce an OSSF lepton
pair, simply by conservation of quantum numbers, it is clear that spectra with sneutrino
NLSP allow for ${\mathcal O}(1)$ probability for production of additional leptons.

In the simplest case, a neutralino sits between the RH and LH sleptons.  The RH
sleptons can then decay via emission of this neutralino and a lepton, assuming 
the neutralino has non-negligible Bino fraction.  The neutralino subsequently decays 
into LH sleptons or sneutrinos, potentially generating one or more additional leptons.
Below, we will consider this as one of our standard decay scenarios, but there are other 
possibilities.

In the cases where all charginos and neutralinos are more massive than the RH and LH
sleptons, decays of the RH slepton proceeds virtually.  The rates for these decays
generally depend in detail on the mass spectrum and mixing matrix for the neutralinos.
For example, it is possible for a heavy Bino-like neutralino to contribute more than
a lighter Wino-like neutralino, or for different levels of neutralinos to contribute
comparably and interfere.  We will assume here that a single mostly-gauge eigenstate
dominates, in which case the relative branching fractions into the different flavors of
LH sleptons and sneutrinos are equal up to phase space.

\begin{figure}[t]
 \centering 
\includegraphics[width=6.4in]{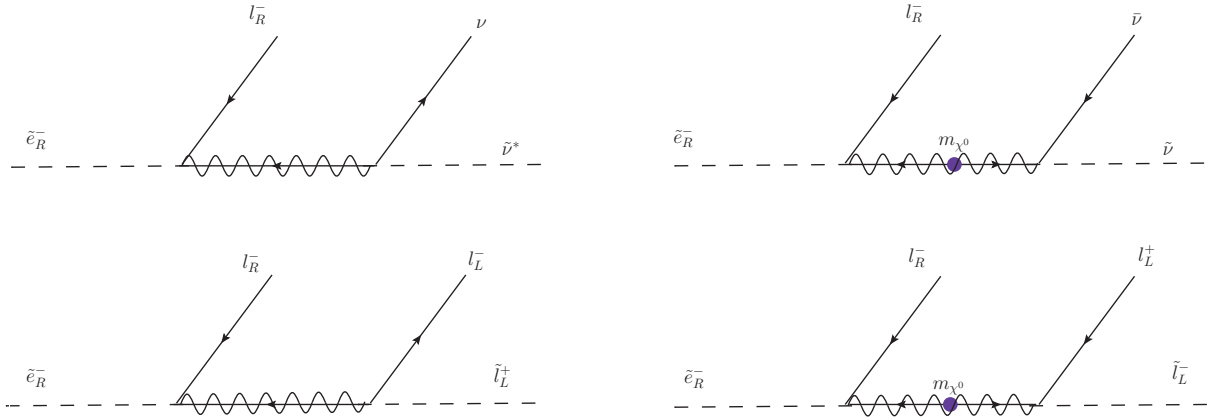}
\caption{Decay modes of the RH sleptons through off-shell neutralinos.  Note that 
the rates from diagrams on the right side are enhanced by 
${\mathcal O}(m_{\gaugino}^2/m_{\sel}^2)$ due to the mass insertions of the neutralino.}
\label{fig:3}
\end{figure}

We show the diagrams for the different 3-body decay modes in Fig.~\ref{fig:3}.  In terms
of charge and flavor flows, these are no different from the equivalent processes
proceeding through an on-shell neutralino.  But besides the obvious changes in kinematics, 
there will also be a bias in charge.  While an on-shell neutralino would decay with equal
rate to $\slep_L^+l_L^-$ and $\slep_L^-l_L^+$ (alternately $\snu^*\nu$ and $\snu\bar\nu$), 
the off-shell neutralino propagator is sensitive to chirality flow.  In particular, processes
with a chirality flip are enhanced by ${\mathcal O}(m_{\gaugino}^2/m_{\sel}^2)$ compared to
those without the flip, favoring decays into opposite-sign leptons.  This would
be an interesting effect to observe, but may be quite difficult in practice.\footnote{A very
similar effect in minimal gauge mediation was pointed out in~\cite{Ambrosanio:1997bq}. In that
case, decays of RH sleptons into the RH stau NLSP are also sensitive to the chirality flow.}  
The production
and decay of the RH slepton (via these neutralino-exchange modes) will always lead to an 
OSSF pair of leptons,  and it will not be obvious which of these two came from 
the RH slepton's {\it decay}.

The overall magnitudes of the 3-body decay widths are naively unimportant for phenomenology, as
long as the decays are prompt.  
However, we need to know these in order to determine the branching fractions for the
flavor non-universal decays, to be discussed shortly.\footnote{Direct decays to gravitinos are also
technically possible.  However, it is quite non-generic for these to be important except in cases
of accidental mass degeneracies or very heavy gauginos.}  
Formulas for the partial decay widths can be found in the appendix.

\subsection{Right-Handed Sleptons: Flavor Non-Universal Decays from Left-Right Mixing}\label{lrmixing}

A more complete treatment of the decays of RH sleptons includes modes induced by
left-right mixing.  Naively, these are only important for staus, but we will show
below that one can easily find spectra where they are also important for smuons.
This can lead to highly flavor non-universal signals with respect to electrons and muons.

To start, we assume that any effects from flavor violation in the soft terms are
completely absent.  At the input scale, the slepton mass-squared matrices are 
proportional to the unit matrix, and $A$-terms are zero.  These will develop
flavor non-universal contributions from running, originating from the Yukawas,
but their presence will not qualitatively change the picture.  For simplicity,
we assume that they are subdominant to the tree-level Yukawa effects.\footnote{This
is the usual situation for the $A$-terms when $\tan\beta$ is large.  We also note that
there may be lepton flavor-violating contributions from 
gravity-scale mediation effects, or from running through
the see-saw threshold.  We further assume that these are small.}  This situation
naturally holds in low-scale mediation scenarios, such as (general) gauge mediation.

Given these assumptions, the dominant flavor effects in the soft masses are
the left-right mixing terms induced by ($F$-term) Yukawa couplings to the Higgs VEVs. 
Each generation of sleptons has an independent $2\times 2$ matrix of soft
masses, of the form
\beq~\label{mixmatrix}
M^2 = \left( \begin{array}{cc}
       m_{\slep_L}^2 & -\mu v Y_l \sin \beta \\
       -\mu v Y_l \sin \beta  & m^2_{\sel}                    
\end{array} \right)~. 
\eeq
Here $v$ denotes the VEV of the Higgs, and $Y_l$ is the SM Yukawa coupling for 
lepton $l$, defined as $Y_l = m_l/(v\cos\beta)$.  (We work in a basis where $\mu$ is real.)  
The mixing term is small compared to
the soft masses, leading to a small left-right mixing angle, as long as the LH and RH
sleptons are not very degenerate:
\beq
\delta_l \simeq \frac{\mu v Y_l \sin \beta}{m_{\sel}^2-m_{\slep_L}^2} = 
\frac{\mu m_l \tan \beta}{m_{\sel}^2-m_{\slep_L}^2}~.
\eeq   

\begin{figure}[t]
 \centering 
\includegraphics[width=6.4in]{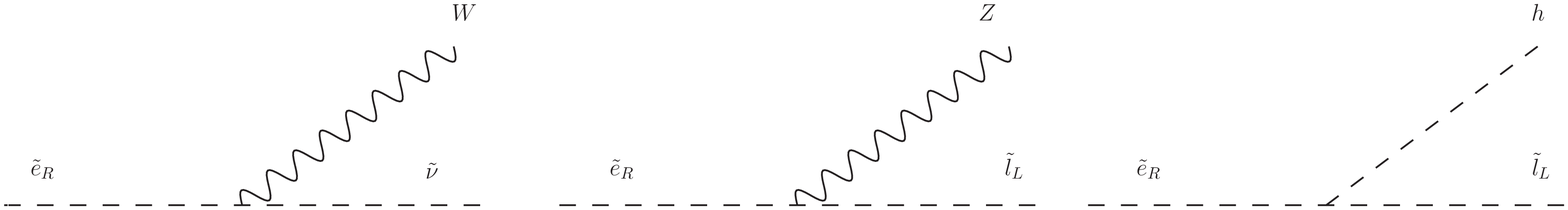}
\caption{Possible two-body electroweak decays modes of the RH sleptons.}
\label{fig:4}
\end{figure}

Usually, the effects of left-right mixing in the first two generations do not have
a significant impact on collider signatures.  However, we saw cases above where the
RH sleptons were forced to decay down to the LH slepton doublet via 3-body processes
mediated by off-shell gauginos.
With left-right mixing, a RH slepton can interact with electroweak gauge bosons, opening 
up additional 2-body decays if  
$m_{\sel}-m_{\slep_L} \simgt 100$~GeV (Fig.~\ref{fig:4}).  It is not difficult to
find spectra where these 2-body, flavor-dependent decays become very important for
smuons.  However, it is practically impossible for them to be relevant for selectrons,
since the Yukawa is too small.  This mismatch in the behavior of RH smuons versus RH
selectrons will ultimately manifest at detector level as an asymmetry between
muons and electrons.  While the presence of RH selectrons in a chain essentially guarantees 
the production of an OSSF $e^+e^-$ pair, the muon-number from RH smuons may 
``disappear'' into sneutrinos.  RH smuon decays will sometimes generate 
additional leptons, from $W$ or $Z$ decay, but these will be flavor-uncorrelated.

The rate for the decay of a RH slepton directly into a sneutrino via on-shell $W$ emission
is
\beqarray\label{seltoWsnu}
\Gamma(\sel \to W \snu) & = & \frac{\alpha_2 \delta_l^2}{8} 
\frac{\lambda^{3/2}(m^2_{\sel}, m^2_{\snu}, m^2_W)}{m^3_{\sel} m^2_W} \nonumber \\
    & = & \frac{Y_l^2}{16\pi} \frac{\mu^2\sin^2\beta}{(m_{\sel}^2-m_{\slep_L}^2)^2} 
\frac{\lambda^{3/2}(m^2_{\sel}, m^2_{\snu}, m^2_W)}{m^3_{\sel}} 
\eeqarray
 with 
\beq 
\lambda (x, y, z) \equiv x^2 +y^2 +z^2 -2xy-2yz-2xz~. 
\eeq
By angular momentum conservation, the decays are exclusively into longitudinal $W$,
which effectively couples as a Goldstone boson.  The analogous decay into $Z\slep_L$
is simply two times smaller than this, with the kinematic factor $\lambda$ appropriately
modified.

There is also a 2-body decay mode into a physical Higgs, with parametrically very similar rate,
again controlled by Yukawas.  The rate is
\beq\label{seltohl}
\Gamma(\sel \to h \slep_L) = \frac{Y_l^2}{32\pi} \, \mu^2 \cos^2\alpha \,
\frac{\lambda^{1/2}(m^2_{\sel}, m^2_{\slep_L}, m^2_h)}{m^3_{\sel}}~,
\eeq
where $\alpha$ is the physical Higgs mixing angle, which tends to be close to zero
for large $\tan \beta$.

These formulas should be compared to those for the 3-body RH slepton decays, which
can be found in the appendix.  The full parametric dependences of the relative rates
are rather involved, even in kinematically simplified limits.  However, we can
perform a comparison of the mass-independent factors, with the understanding
that there is still room for a large amount of numerical engineering.  We find
\beqarray
\frac{\Gamma_{\rm 2-body}}{\Gamma_{\rm 3-body}} & \sim & \frac{Y_l^2\sin^2\beta/16\pi}{\alpha_1^2/2^5\pi} 
\times ({\rm mass \; factors}) \nonumber \\
  & \sim & \left\{ \begin{array}{l} 2\times10^{-4} \\ 8 \\ 2400 \end{array}  \right\}   
           \left(\frac{\tan\beta}{10}\right)^2 \times ({\rm mass \; factors})~
\eeqarray
for ($e$,$\mu$,$\tau$), respectively.  It is clear that RH selectrons will practically never
decay via 2-body, whereas RH staus will very likely be dominated by 2-body, when those modes
are kinematically available.  The RH smuon occupies a highly sensitive point, such that the
relative 2-body rate will depend on the detailed mass spectrum and $\tan\beta$.  Below, 
we will investigate
spectra where this relative rate is both large and small.

Left-right mixing effects can also be important for more general sets of spectra, again tending
to hide the muon-number in RH smuon decays.  For example,
we can have $m_{\sel}-m_{\slep_L} \simlt m_W$, such that 2-body electroweak decays are shut off.  
However, 3-body electroweak decays will still contribute.  If the mostly-Bino neutralino is 
relatively heavy and unmixed, then decays via left-right mixing may nonetheless be preferred for smuons.  
Another possibility
is a mostly-Wino in between the RH and LH sleptons.  If the neutralino mixing is small enough,
RH smuons may again prefer to decay via left-right mixing.  Decays into the mostly-Wino neutralino 
will then proceed anyway because of {\it this} mixing, but so will decays (with twice the rate) 
to charginos, and to gauge bosons
if the phase space is available.  We will not analyze these cases in detail, as they will
lead to qualitatively the same effects  as cases where decays into on-shell electroweak 
bosons dominate.

\section{Leptonic Signals}\label{Signals}

We now describe our main signals in detail, before moving on to collider studies
in section~\ref{Simulations}.  We will first discuss the 
generic multileptonic signals of RH-active spectra.  We will then focus on the signals with 
the highest rates and
the least combinatoric ambiguities, namely dileptonic and trileptonic channels.
Subsequently, we will see how flavor non-universalities can manifest in these channels,
if electroweak decays of the RH smuons are important.  Finally, we will discuss the
tau-independent signals of the NUHM spectra, and how the techniques for investigating
RH-active spectra can also apply in this case.

\subsection{High-Multiplicity Signals of RH-Active Spectra}

As we have seen in section~\ref{decays}, chains with a RH slepton can potentially
generate a large number of leptons.  The most extreme case would be 
$\gaugino^0 \to l\sel \to l(ll\slep_L) \to lll(l\nu\snu)$.  If this were to
occur on both sides of an event, the number of leptons would tally to eight.
Of course, this exceptional class of events is also quite rare, since many branching
fraction penalties and detection efficiencies would have to be paid.  
But this still tells us that we can expect these events to be quite ``leptophilic.''

Such behavior contrasts with that of spectra where RH sleptons do 
not significantly participate, where we were relatively lucky to get an observable
4-lepton signal after a 100~fb$^{-1}$ LHC run.  There, neither the production nor the 
decay of $\slep_L$ was
guaranteed to produce a lepton.  We nonetheless could expect to find
significant dilepton and trilepton signals.  With spectra where RH sleptons are
produced with an appreciable rate, we will see
below that we
might reasonably find observable multilepton rates up to 5-lepton or 6-lepton, with healthy populations
in the lower-multiplicity bins.  Needless to say, the
backgrounds for such dramatic signals are small.

Similar behavior was pointed out in~\cite{DeSimone:2009ws} in the 
context of low-scale gaugino mediation.  This is not surprising, given that 
the spectra which we analyze here are in some cases nearly the same, but 
with the ordering of the LH and RH sleptons reversed.  However, we point out that
there will always be significant differences in the phenomenology of these spectra.
The specific scenario investigated in~\cite{DeSimone:2009ws} contained a metastable
mostly-RH stau as the NLSP, leading to striking CHAMP signals.  
Even if the 
SUSY-breaking scale were lowered, such that RH sleptons decay promptly, then every
chain is guaranteed to independently produce an OSSF lepton pair or tau pair.  Such
a scenario would be much more leptophilic than ours.  It is also possible that RH
selectrons and smuons first decay to stau, which then decays to tau and gravitino,
leading to at least four taus per event, in {\it addition} to the two OSSF 
pairs.  If even a fraction of these taus and leptons are detected with good efficiency
for every event, they will still be quite suggestive of scenarios with RH slepton NLSP.

Although high-multiplicity leptonic channels will serve as very clean evidence for new physics, 
we will not utilize signals with four or more leptons
for any analysis beyond simple counting.  The main reasons for this are the formidable 
combinatorial uncertainties and the overall lower statistics.  In any case,
the only new kinematic information contained in these events would be the 4-lepton
distributions from a single decay chain.  For the purposes of this study, we
consider these distributions as lost, and
instead we will concentrate on the dileptonic and trileptonic events.

\subsection{Dileptons}\label{2lsignals}

\begin{figure}[t]
 \centering 
\includegraphics[width=3.5in]{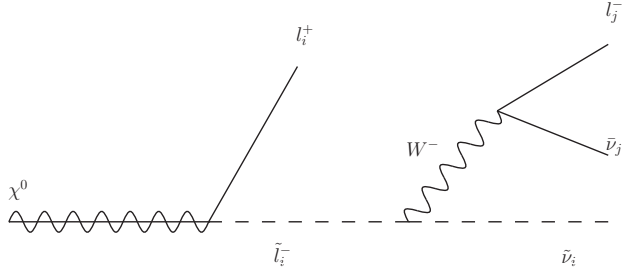}
\caption{The sub-cascade responsible for the OS dilepton excess in chains with LH sleptons.}
\label{fig:6}
\end{figure}

In~\cite{Katz:2009qx}, the dileptonic distributions had two major classes of contributions.
The first was from events where one chain produced opposite-sign, flavor-uncorrelated leptons
from LH slepton production and decay (Fig.~\ref{fig:6}), and the other chain 
produced no visible 
leptons.  The second contribution came from events where each chain produced a single lepton
(from either slepton production {\it or} decay).  These leptons were totally uncorrelated
in both sign and flavor.  The full set of SUSY dilepton invariant mass spectra contained a 
broad contribution from this latter class of events, equally distributed between all sign 
and flavor bins.
The correlated leptons from a single chain, on the other hand, led to a more localized
bump, which appeared with equal rates in OSSF and OSOF channels.
The correlated distribution could be extracted using a simple sign-subtraction
procedure, and its peak could be used to infer relationships between the gaugino, LH
slepton, and sneutrino masses:
\beq
m_{ll}^{\rm peak} \approx 0.48\ 
\sqrt{\frac{(m_{\gaugino}^2-m_{\slep}^2)(m_{\slep}^2-m_{\snu}^2)}{m_{\slep}^2}}~.
\eeq 

\begin{figure}[t]
 \centering 
\includegraphics[width=3.5in]{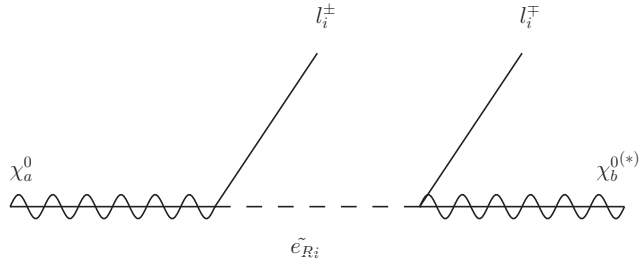}
\caption{The sub-cascade responsible for the OSSF dilepton excess in chains with RH sleptons. 
Neutralino $\chi_b^0$ can
be either on- or off-shell, and in the latter case we usually have $a=b$.}
\label{fig:5}
\end{figure}

\begin{figure}[t]
 \centering 
\includegraphics[width=5.5in]{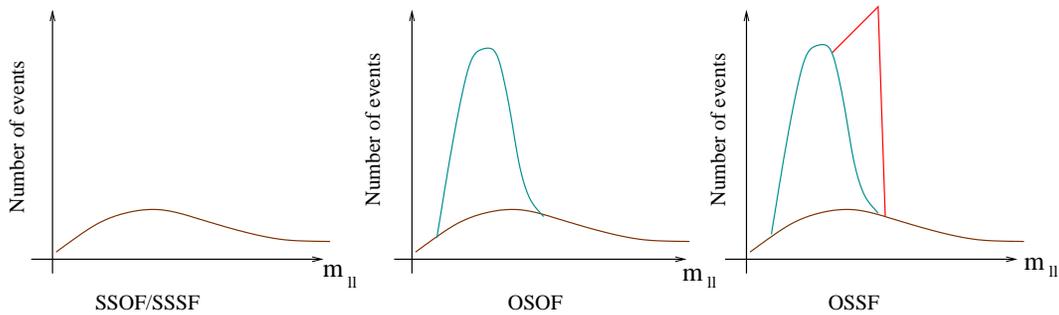}
\caption{Illustration of the dilepton invariant mass distributions characteristic
of RH-active spectra, broken down into sign/flavor channels.  The independent contributions from
LH sub-cascades, RH sub-cascades, and uncorrelated dileptons are shown in teal, red, and brown, 
respectively. }
\label{fig:distrib}
\end{figure}

These signals will persist in RH-active spectra, but there will also be one qualitatively
new contribution.
The sub-cascade $\chi^0_a \to \sel \to \chi^{0(*)}_b$ (see Fig.~\ref{fig:5}) produces
two leptons correlated in both sign and flavor.  When these are the only leptons produced
in a chain, and the other chain in the event produces no leptons, a new dilepton excess
is generated in the OSSF channel.  This signature is well-known in traditional
spectra with a neutralino LSP, where it is separated from the uncorrelated dilepton distribution
by using a flavor-subtraction, OSSF-OSOF.  In sneutrino NLSP spectra, this signal will coexist 
with the OS distribution from LH slepton production/decay, as well as the 
background of totally uncorrelated SUSY dileptons.  The situation is illustrated in 
Fig.~\ref{fig:distrib}.

The shape of the OSSF excess depends on further details of the decay
process.  In the case that a neutralino sits below the RH slepton, such that the 
RH slepton can undergo a 2-body decay, the distribution will be of the 
characteristic ramp-and-edge shape.  If the decays are 3-body, then the 
distribution takes on a more bump-like shape, similar to the shape of the OS 
excess.  However, the distribution will usually be skewed to some extent toward 
higher masses.

Clearly, as Fig.~\ref{fig:distrib} suggests, extracting both of the independent
kinematic shapes from the different sign and flavor channels is slightly nontrivial,
but still straightforward.
We should simply perform two independent subtractions:  OSSF-OSOF to
reveal the RH contribution, and OSOF-SSOF to reveal the LH
contribution.\footnote{It is possible (though not generic) to get 
sign correlations in the contributions
from leptons produced in independent decay chains, due to biases in production.
In that case, the shapes in the OS and SS channels will likely be very similar, but
their normalizations could be different.  It would then still be possible to perform
a weighted OSOF-SSOF subtraction, such that the high-mass tail is canceled off.}
Unfortunately, it would be impossible at this level of analysis to independently
extract the LH contribution within the OSSF channel, and verify that it is equal
to the contribution within the OSOF channel.  However, the crucial observation
at this point, already suggestive of the presence of sneutrinos in the chains, is
that the OSOF-SSOF subtraction gives a nonzero result at all.  More realistically, we 
will require that this subtraction gives a statistically significant excess above
opposite-sign, flavor-uncorrelated backgrounds from the Standard Model, such as $t\bar t$.
We will demonstrate that this is possible in section~\ref{Simulations}.

More generally, it is possible that the OSSF contribution from LH sleptons has
a different normalization from those in the OSOF channel, for example due to the
interference effects mentioned in subsection~\ref{lhdecays}.  While we have just
emphasized that this cannot be independently checked, we note that with high 
enough statistics it may be possible to see additional peaks or dips in the shape 
of the OSSF-OSOF excess.  The presence or absence of such features becomes particularly 
informative if the OSSF-OSOF and OSOF-SSOF shapes are sufficiently distinct.

\subsection{Trileptons}\label{3lsignals}

The trileptonic channel will also have a sizable signal, more so than in the
case without the RH sleptons.  Originally, the main way to obtain trilepton
was to have one chain produce two leptons from LH slepton production and decay
(Fig.~\ref{fig:6}), and 
the other chain to produce one additional lepton.  The combinatorial ambiguities
were therefore fairly simple, and we demonstrated in~\cite{Katz:2009qx} a straightforward
way to identify which two leptons were produced in the same chain.  Now, this channel
can have a variety of contributions from the production and decay stages of both
RH and LH sleptons, including the option for all three leptons being produced in the
same chain.

Let us start analyzing the possibilities by categorizing the sign and flavor
content of the trileptons.  Generally, the majority of these events will have two 
same-signed leptons and a single ``uniquely-signed'' lepton.  In other words, trileptonic
events with all leptons of the same sign will be subdominant.  For example, if the
leptons were all totally uncorrelated, only 1/4 of the events would be fully same-signed.
More realistically, this fraction will be even smaller, since there are often physical
sign correlations, most obviously in the OS dileptons produced in either RH or LH
production/decay.  We will subsequently focus our analysis exclusively on the cases with a 
uniquely-signed lepton, since we do not expect all-same-signed to carry much useful
kinematic information.  However, the observation of same-signed trilepton events
is yet another clue to the simultaneous presence of RH and LH sleptons, and may serve
as an extremely clean (re-)discovery signal.

Given that we have one uniquely-signed lepton, we can further classify the flavor
structure of these events.  We compare the flavor of the uniquely-signed lepton
with the flavors of the two same-signed leptons, leading to three distinct cases:
\begin{itemize}
 \item SFSF - both parings are of the same flavor, those are either all-electron
or all-muon events,
 \item SFOF - one pair is of the same flavor and the second pair is of the opposite flavor,
 \item OFOF - both pairings include opposite flavors.
\end{itemize}
We can now perform counting experiments within these three channels, to attempt to 
deduce the composition of the events.

First, consider spectra with relatively inactive RH sleptons, where all trilepton events
contain a LH slepton production/decay.  In this case, while
two of the leptons are sign-anticorrelated, the flavors of all three leptons are
completely uncorrelated.  For example, if the uniquely-signed lepton is an $e^-$,
then there is equal probability for the same-signed pair to be
$e^+e^+$, $e^+\mu^+$, $\mu^+e^+$, and $\mu^+\mu^+$.\footnote{Note that  $e^+\mu^+$ and
 $\mu^+e^+$ are indeed distinct.  The general event structure is $l_i^-l_j^+l_k^+$, with
each of the three leptons ($i$,$j$,$k$) produced in independent, flavor-blind subprocesses.}  
This correspond to SFSF, SFOF, SFOF, and OFOF, respectively.  Quite generally, then, we
expect the counting ratio OFOF:SFOF:SFSF to be 1:2:1.

Now suppose instead that the trilepton contribution is exclusively
from RH slepton production/decay in one chain, with additional uncorrelated lepton,
either emitted in the other chain or in a subsequent LH slepton production or decay.  
The first process is guaranteed to produce an OSSF pair, and one of
these is in turn guaranteed to be the uniquely-signed lepton in the event.  Given
that one of the remaining leptons is perfectly flavor-correlated with this, and the
other perfectly flavor-uncorrelated, we expect equal contributions to SFOF and SFSF, 
and vanishing OFOF.  We therefore get OFOF:SFOF:SFSF of 0:1:1.

RH-active sneutrino NLSP spectra will exhibit a superposition of 1:2:1 and
0:1:1, with the latter represented in proportion to the amount of RH slepton
production.  The observation of both of these contributions added together would serve
as a powerful supplement to the interpretation already suggested by the dilepton
analysis.  The simplest way to check this is to first verify the presence of a 
substantial OFOF contribution, and then to add together OFOF and SFSF, and
see if the sum matches SFOF.

There will also be additional flavor-uncorrelated contributions,
even from chains with RH sleptons.  For example, we may fail to reconstruct
one of the leptons from its production or decay, and pick up two other unrelated
leptons.  However, these events do not tend to be very common.  In spectra with LH and
RH together, on the other hand, the flavor-uncorrelated events would usually constitute
a major fraction of the trileptons.  A simple indicator of the presence of the sneutrino
NLSP is therefore the relative size of the OFOF, which is 
nominally absent or small for trilepton signals dominated by RH.  Equivalently, we
should pay attention to the fraction of trilepton events that {\it lack} an OSSF
pairing.

Having established the basic trilepton sign/flavor channels, we can now proceed to 
investigate kinematic distributions.  We can first attempt to recover the analysis
of~\cite{Katz:2009qx}, in which we ``rediscover'' the dilepton invariant mass
distribution from LH slepton production and decay (now extracted via OSOF-SSOF 
subtraction, as above).  The easiest way to do this
in the presence of RH contamination is to focus on the OFOF channel.  This reduces
statistics by a factor of four, but leads to a signal which is much easier to 
interpret and compare to the dileptons.

We can further inquire into whether we can extract kinematic distributions involving the 
RH sleptons, which will occupy the SFOF and SFSF channels.
There will be two 
main classes of RH events contributing, depending on where the third lepton comes from:
either it is produced along with a LH slepton further down the chain, or it comes from the other 
side of the event.\footnote{It may also come from the {\it decay} of 
the LH slepton further down the chain.  This would require us to miss the lepton associated
with the LH slepton production (lest the event end up classified as 4-lepton), as well as to pay 
the branching fraction price of 22\%.  These therefore represent a subdominant contribution.}  
Since the flavor of the third lepton is uncorrelated in both cases, we cannot 
use this information to disentangle the two distributions.
There is potentially a charge bias for leptons produced along with the LH sleptons in 3-body RH slepton
decay, as pointed out in subsection~\ref{rhdecays}.  However, we also pointed out that it might be 
very difficult to spot.  We therefore tentatively consider these two samples to be
completely entangled.

This must then be added to the LH contribution.  But, in principle, we can subtract out this
contribution by utilizing its different flavor structure.  OFOF is dominantly LH, and we 
can predict its contribution to the other channels using the 1:2:1 ratio above.

In particular, we can access the OSSF dilepton distribution embedded in the trilepton sample
as follows.  First, we construct the LH-dominated dilepton distribution in OFOF by making 
{\it random} pairings with the
uniquely-signed lepton.  This will contain equal amounts of correct pairings and incorrect
pairings.  Then we form the dilepton distribution in SFOF, using the unique
OSSF pair.  This will contain the distribution we are after, as well as the LH contamination,
also randomly paired due to the lack of flavor structure.
Last, we subtract off {\it twice} the distribution of OFOF.  Assuming we have enough statistics
to obtain a meaningful final distribution, this can be checked against the OSSF excess 
found in the dileptonic channels.

Using similar logic, we can try to access the trilepton invariant mass distribution from RH slepton
production and decay, with subsequent LH production in the same chain.   The precise shape 
of this distribution with all of the intermediate states on-shell was studied in detail 
in~\cite{Miller:2005zp}.  It can serve as a useful supplement to the kinematic information
already available from the dilepton distributions.  To maximize statistics, we take all 
trileptons from the SFOF and SFSF channels (equivalently, all events in which we can find
an OSSF pair), and subtract from this {\it three} times the trilepton distribution of the
OFOF events.  This will leave over the contribution of genuine correlated trileptons,
all from the same chain, but combined inextricably with the uncorrelated RH distribution.
Of course, even if this subtraction is too awkward to carry out due to limited statistics, 
or the purification ends up being only modest due to the unsubtractable RH combinatorial background, 
we may still simply plot the SFOF+SFSF distribution and look for a bump.  For the samples which
we study in section~\ref{Simulations}, at 100~fb$^{-1}$ luminosity, we take the latter tactic.

\subsection{Flavor Non-Universal Signals from Left-Right Mixing}\label{flvor-bias}

As we discussed in subsection~\ref{lrmixing}, it is possible for otherwise flavor-blind
spectra to exhibit large differences between the decays of RH selectrons and RH smuons.
The latter can have non-negligible left-right mixing originating from the Yukawa couplings.
This can cause the RH smuons to 
dominantly decay via electroweak
or Higgs boson emission, and ultimately lose their muon-number to a sneutrino.  The RH
selectron, on the other hand, would usually decay via neutralino emission, converting into
an electron.

We can immediately infer the effects of these flavor non-universal decays on the dilepton signals.
The OSSF excess that would otherwise have been generated in RH smuon production and decay
will no longer exist.  In principle, then, we could observe that the excess in OSSF is entirely
composed of electron pairs.  Of course, the situation is not so simple, since we do not know
which pairs came from RH sleptons event-by-event.  But we can still isolate the presence
of the excess by subtracting OSSF($e^+e^-$)-OSSF($\mu^+\mu^-$).  This should practically equal the
OSSF-OSOF distribution.  A similar strategy was employed 
in~\cite{Kumar:2009sf}, for spectra with lepton flavor violation in the SUSY soft terms 
(incorporating RH sneutrinos).

The RH smuon will not completely disappear from the dilepton channels, but its signatures 
will be highly
attenuated and redistributed.  Decays mostly proceed into $W^{(*)}\snu$, with the RH smuon acting like
a LH smuon with a larger mass gap.  In these decays, we get the usual 22\% chance to generate
a second lepton of opposite-sign but uncorrelated-flavor.  We will therefore get a small enhancement of
the OS excess, albeit at higher invariant mass than that caused by genuine LH sleptons.  It will also
be flavor-biased, consisting only of $\mu\mu$ and $\mu e$.  Decays into $Z^{(*)}\slep_L$ and $h^{(*)}\slep_L$ can
further contribute when the $\slep_L$ decays leptonically.\footnote{A small fraction will also contribute
to trilepton, when the $Z$ decays leptonically.}  Finding these signals may be relatively difficult 
without rather high statistics.  To first approximation, we will consider the dileptonic signals 
of the RH smuons as lost.

The deficit of RH smuon decays will also manifest in the trilepton channels.  The channels
containing an OSSF pairing (SFOF+SFSF) will generally contain fewer muons.  One way
to see the effect is to simply plot the number of muons in these events.  If the signals
were perfectly flavor universal, the trilepton signal from RH sleptons would contain
$(0\mu,1\mu,2\mu,3\mu)$ in equal ratio.  The signal from LH sleptons (with no flavor structure
whatsoever) gets added in with
ratio 1:3:3:1.  We therefore expect, quite robustly, that completely flavor-blind chains will
have equal amounts of $0\mu$ and $3\mu$, and equal amounts of $1\mu$ and $2\mu$.
In the spectra with significant left-right mixing,  the higher multiplicities are depleted, leading
to biases $0\mu > 3\mu$ and $1\mu > 2\mu$.

Of course, it is also possible to have an intermediate case, where some non-dominant fraction of 
RH smuon decays go through 
electroweak channels.  This would lead to smaller, but possibly still observable flavor biases.

We expect that similar kinds of electron-muon asymmetries can manifest in leptogenic 
SUSY~\cite{DeSimone:2009ws}.  Indeed, the possible importance of left-right mixing on 
(LH) smuon decays was also pointed out in that context, though there the muon counting was 
highly ``contaminated'' by misidentified stau CHAMPs.  In similar scenarios with promptly-decaying 
stau NLSP, it may be possible to immediately observe an asymmetry between electrons and
genuine muons.  It would be interesting to study these effects in spectra similar 
to~\cite{DeSimone:2008gm, DeSimone:2009ws} in more detail.

\subsection{Signals with a Light Stau Doublet}\label{NUHMsignal}

\begin{figure}[t]
\centering 
\includegraphics[width=5.9in]{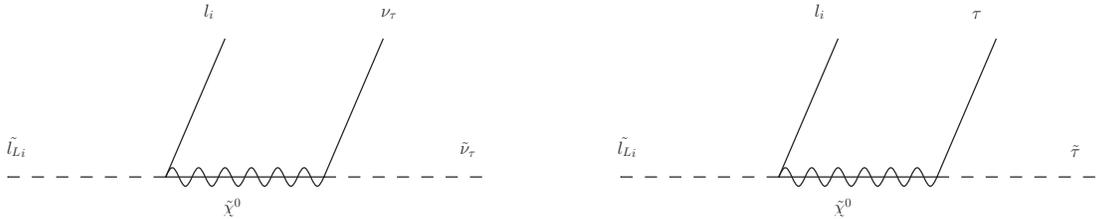}
\caption{Important decay modes of the LH slepton in spectra with a light stau doublet. 
Note that the lepton flavor is correlated with its parent.}
\label{fig:7}
\end{figure}

If the mediation scale is relatively high, the LH stau doublet can be significantly pushed down
in mass compared to the first two generations, due to Yukawa effects in the running
of the soft masses.  In particular, this occurs in the NUHM spectra of~\cite{Ellis:2002iu, Ellis:2008as}.  
We saw above, in subsection~\ref{lhdecays}, that splitting off the third generation can 
significantly change the decay modes of the LH sleptons of the first two generations, introducing
sizable branching fractions into $l(\snu_\tau \nu_\tau)$ and $l(\stau\tau)$.\footnote{Detailed
analytic expressions can be found in~\cite{Kraml:2007sx}.  Expressions for the case of a single
neutralino dominating the decay can be found in our appendix, though we note that interference
effects between Bino and Wino can be quite substantial, as the masses and couplings are in direct
proportion in high-scale scenarios with unified gaugino mass.  Decays
into $\nu(\snu_\tau\tau)$ and $\nu(\stau\nu_\tau)$, via chargino exchange, will also be present,
but they do not significantly change the phenomenology.}   The relevant diagrams
are illustrated in Fig.~\ref{fig:7}.

When dealing with these spectra, we shall revert to the assumption that RH sleptons are bypassed.
Nonetheless, in some sense the LH sleptons of the first two generations take their place.  The
new decay modes now serve to populate an excess in the OSSF dilepton channel, in addition
to the OS excess from the competing $W^*$ decay modes.  The presence of these two overlapping
excesses can be inferred, and their distributions separated, exactly as for the RH-active 
scenarios discussed above.

These spectra will also appear to be quite leptophilic.  Tau production is now naturally quite
high, and each tau has an approximately 35\% chance of manifesting as an isolated electron
or muon.  In fact, since tau (and stau) decays proceed through $W^*$, we end up with
several new opportunities for OS, flavor-uncorrelated dilepton signals.  These signals will
usually be biased towards low invariant mass, as the energy of the original tau must be
shared with two neutrinos.  These new OS distributions will be irreversibly added on top of 
the OS excess from the production and $W$-mediated decay of LH sleptons
of the first two generations.\footnote{In principle, we could determine whether a lepton
is prompt, versus a product of tau decay, by measuring the displacement of its track from
the primary event vertex.  However, the distribution of impact parameters is quite broad,
and depends on the unknown energy of the original tau.  Separation of the independent dilepton 
distributions with and without taus may not be feasible without high statistics and
careful analysis.}  
While the presence of this tau ``contamination''
actually increases the leptonic rates, improves chances for discovery of a new physics
signal, and moreover indicates the presence of significant tau production through its
shape, we see that it 
can nonetheless obscure the physics we were originally interested in finding, namely
the presence of light LH sleptons decaying into sneutrinos.  However, we will see
that all of our signals can in principle still be observed, even in the presence
of this new SUSY background.

Naturally, the presence of taus in almost every event could be deduced by applying hadronic
tau tags.  We expect that this will be a clear giveaway of the flavor (but not the charge) 
of the NLSP, even if 
the tag is not very efficient.  However, we will not rely on hadronic taus for detailed
kinematics.  Instead we focus entirely on the performance of our proposed multilepton 
measurements, which directly carry over from our RH-active analysis and will be quite 
robust independent of the hadronic tau efficiency.

\section{Collider simulations} \label{Simulations}

In order to determine what these signals might look like at the LHC, we have performed
simulations of three sample sneutrino NLSP spectra, along with Standard Model backgrounds.
The simulation and analysis methodology is identical to that in~\cite{Katz:2009qx}.
In particular, we generate complete spectra, including radiative corrections, using
{\tt SOFTSUSY v3.0.7}~\cite{Allanach:2001kg}.  We generate SUSY $2\to 2$ pair production using 
{\tt MadGraph/MadEvent v4.3.0}~\cite{Alwall:2007st}. We use {\tt BRIDGE v2.17}~\cite{Meade:2007js} 
to calculate branching fractions and to
simulate the decay chains.\footnote{ We use a modified version of {\tt BRIDGE} 
which incorporates left-right mixing effects for the light generations. 
We are grateful to Matt Reece for his help.}  
We then shower and hadronize with {\tt PYTHIA
v6.4.14}~\cite{pythiamanual}, and perform event reconstruction with {\tt FastJet
v2.3.4}~\cite{Cacciari:2005hq}. 
As before, we do not include detector effects beyond basic geometric
acceptance and simple $p_T$ cuts on  reconstruction.

After hadronization, event reconstruction proceeds as follows.  
We separate out leptons (electrons and muons) with $p_T$
above 5 GeV and $|\eta| < 2.5$,
and check them for isolation.  We scalar-sum the $p_T$ of the lepton with the $p_T$s of 
all other non-leptonic (and non-invisible)
particles within an $\eta$-$\phi$ cone of size $0.4$.  If the lepton constitutes 90\% or more
of the total $p_T$, then we consider it ``tight.''  Failing this, if the $p_T$ of the other particles
tallies to less than 10 GeV, we consider it ``loose.''  (This second class of leptons will 
be used to keep more signal in events with high lepton multiplicity, namely three or more.)  
We set aside leptons which fail both of these criteria for clustering into jets.

After identifying the set of isolated leptons, we proceed to cluster all of the remaining
non-invisible particles
in the event into jets using the Cambridge/Aachen algorithm with $R = 0.4$.  We keep jets with
$p_T > 20$ GeV and $|\eta| < 2.5$.

We focus exclusively on super-QCD production modes, as these are the most spectacular, and
the most straightforward to extract from the backgrounds.\footnote{Electroweak production
of gauginos or RH sleptons may also yield very interesting multilepton signals, which may become
especially relevant for a reduced energy LHC, or for searches at the Tevatron.  We relegate the
question of their observability for future studies.}
We require that each event have at least two jets with $p_T > 300$~GeV, and \met\ $> 200$~GeV.
We study
events with at least one tight lepton, with the further requirement that a second lepton
be tight in events with two or more leptons.  These requirements very efficiently remove leptons from
heavy flavor decay.

Backgrounds are as before, including $t\bar t$, single- and di-boson.  $Z/\gamma$ are treated 
fully off-shell.   We did not investigate new backgrounds relevant for 4-lepton and higher channels, 
as the 3-lepton backgrounds are already essentially negligible given our cuts on jet and \met\ 
activity.\footnote{The dominant 3-lepton background, with about 0.04~fb after cuts, 
is $WZ$+jets.  We have included this in the 
analysis for illustration.  The final sample has very low cross section per Monte Carlo event, and 
appears as a small contribution on 3-lepton plots (with fewer than one event per bin).  $ZZ$+jets
represents an even smaller contribution, and has not been included.}

Since many of our signals will require high luminosity to achieve good statistical
control, we perform our analysis at 100~fb$^{-1}$.  We optimistically assume that the
LHC will be running at the design energy of 14 TeV by this point.  A somewhat
lower final operating energy will not significantly change our conclusions.

\begin{table}[t]
 \centering
\begin{tabular}{|c|c|c|c|}
\hline
   &  RH-active & LR-mixed smuon & NUHM   \\ \hline 
$\gluino$                    &  1403 & 1186 & 1034 \\
$\tilde{u}_L/\tilde{d}_L$    &   935 &  855 &  960 \\ 
$\tilde{u}_R$                &   944 &  878 &  863 \\
$\tilde{d}_R$                &   934 &  861 &  938 \\
$\tilde{t}_1$                &   866 &  801 &  672 \\
$\tilde{t}_2$                &   954 &  875 &  921 \\
$\tilde{b}_1$                &   912 &  824 &  881 \\
$\tilde{b}_2$                &   934 &  860 &  927 \\ \hline
$\gaugino^0_1$               &   285 &  360 &  182 \\
$\gaugino^0_2$               &   386 &  405 &  340 \\
$\gaugino^0_3$               &   458 &  425 &  544 \\
$\gaugino^0_4$               &   516 &  600 &  561 \\
$\gaugino^+_1$               &   384 &  390 &  341 \\
$\gaugino^+_2$               &   515 &  600 &  564 \\ \hline
$\tilde{e}_R$                &   254 &  316 &  408 \\
$\tilde{l}_L^+$              &   199 &  184 &  153 \\
$\tilde{\nu}$                &   184 &  166 &  131 \\
$\tilde{\nu_{\tau}}$         &   182 &  166 &   95 \\
$\tilde{\tau}_1$             &   198 &  174 &  120 \\
$\tilde{\tau}_2$             &   255 &  320 &  386 \\ \hline
\end{tabular}
\caption{Physical masses (in units of GeV) in the three example spectra.}
\label{tab:output_invlep}
\end{table}

\subsection{Simple RH-Active Spectrum}

\begin{figure}[t]
\begin{center}
\epsfxsize=0.49\textwidth\epsfbox{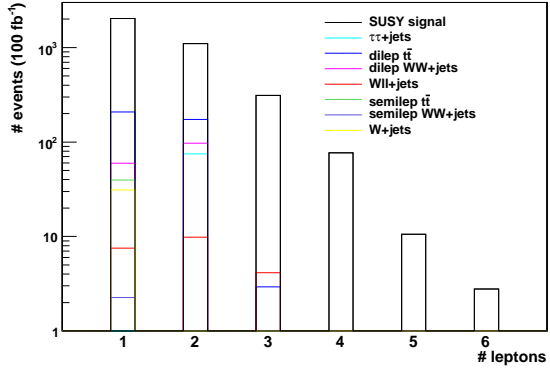}
\caption{(Simple RH-active spectrum.) Multilepton counting, using
a logarithmic scale.}
\label{fig:nl-simple}
\end{center}
\end{figure}

\begin{figure}[t]
\begin{center}
\epsfxsize=0.49\textwidth\epsfbox{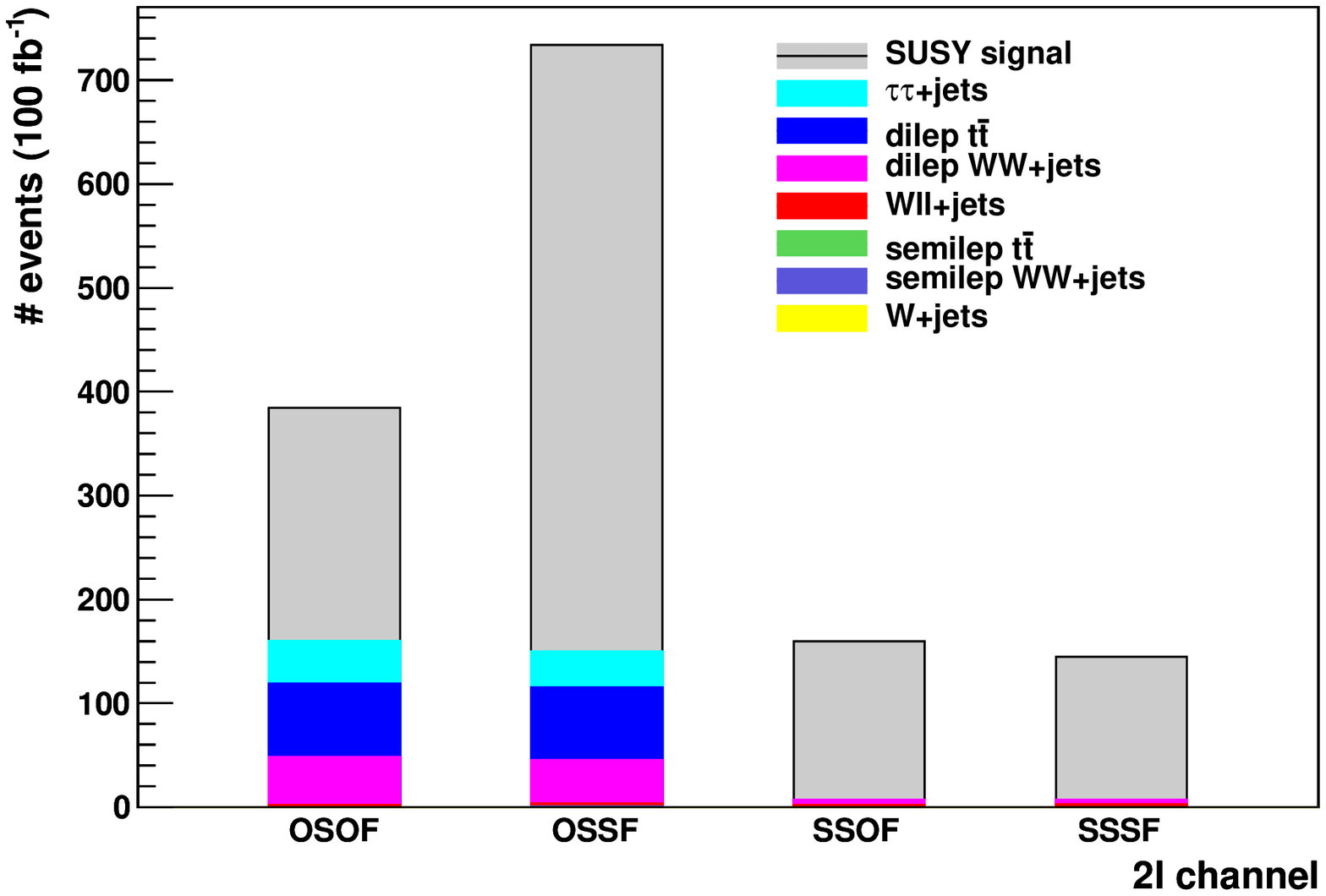}
\epsfxsize=0.49\textwidth\epsfbox{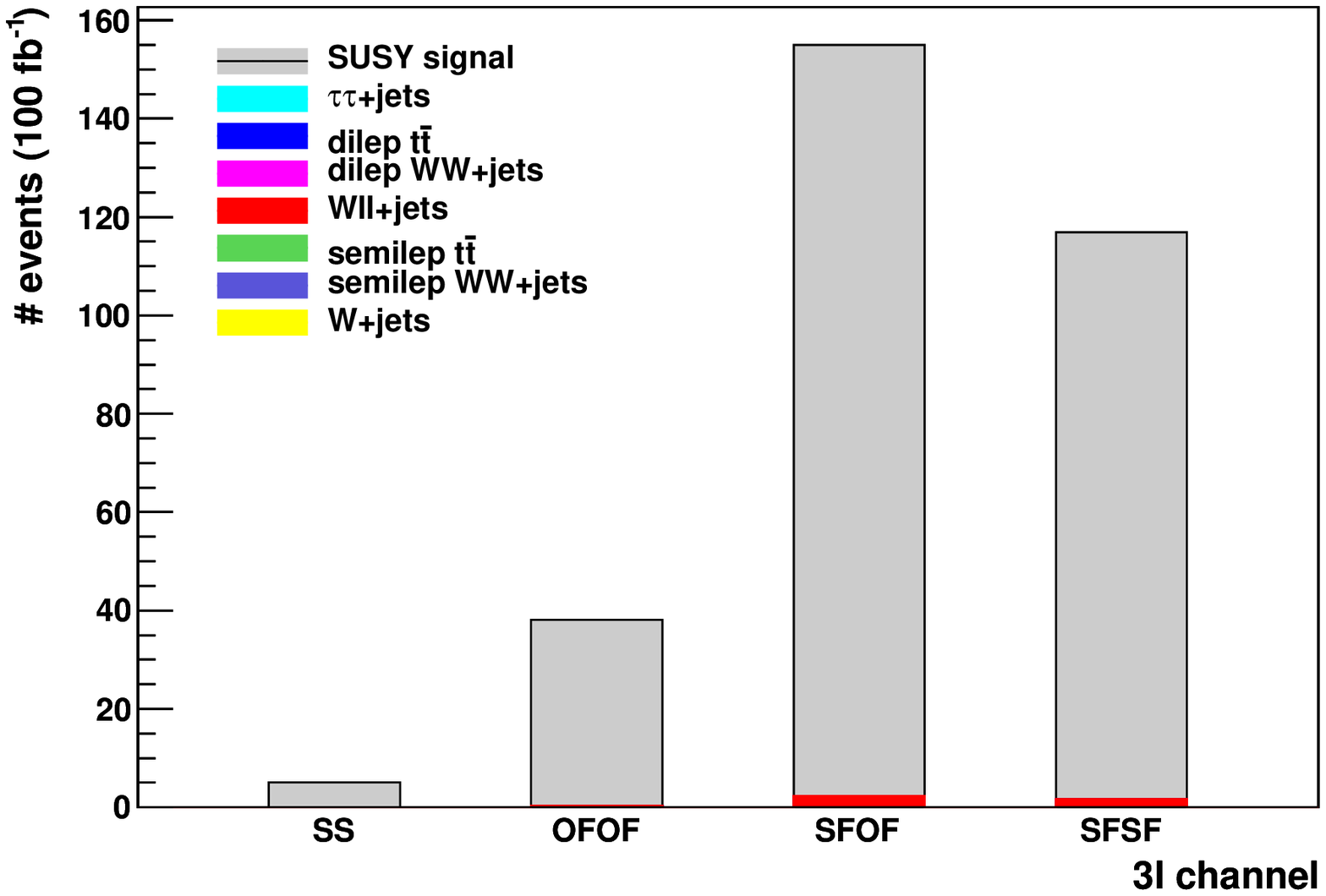}
\caption{(Simple RH-active spectrum.) Relative sign/flavor structure of the dileptonic 
and trileptonic channels.  Histograms are stacked. }
\label{fig:structure1}
\end{center}
\end{figure}


\begin{figure}[t]
\begin{center}
\epsfxsize=0.49\textwidth\epsfbox{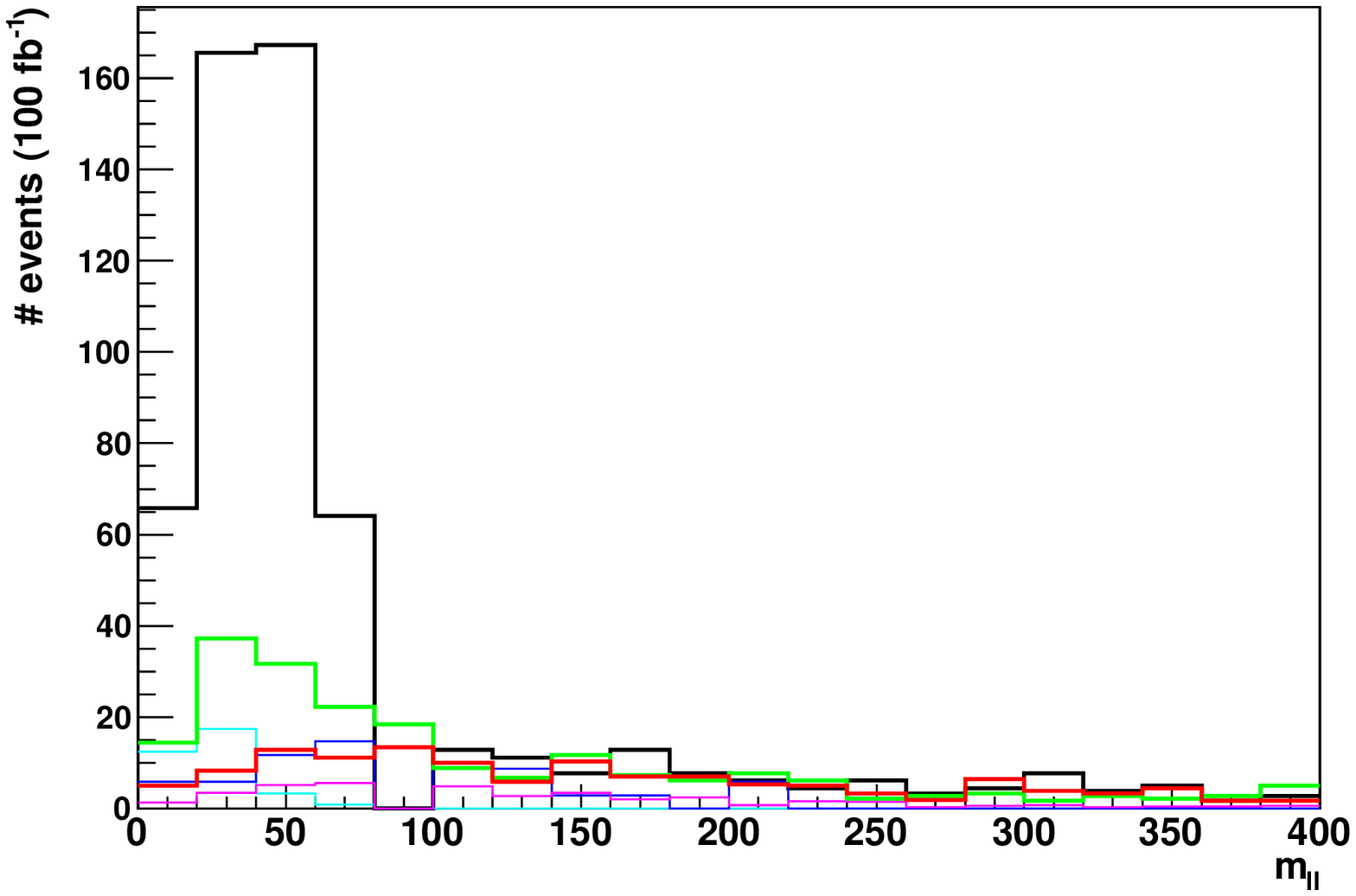}
\caption{(Simple RH-active spectrum.) Dilepton invariant mass distributions.
The signal histograms are OSSF (black), OSOF (green), and averaged SS (red).
The major backgrounds, averaged between OSSF and OSOF, are $\tau\tau$ (cyan), dileptonic
$t\bar t$ (blue), and dileptonic $WW$ (pink).}
\label{fig:mll-simple}
\end{center}
\end{figure}


\begin{figure}[t]
\begin{center}
\epsfxsize=0.49\textwidth\epsfbox{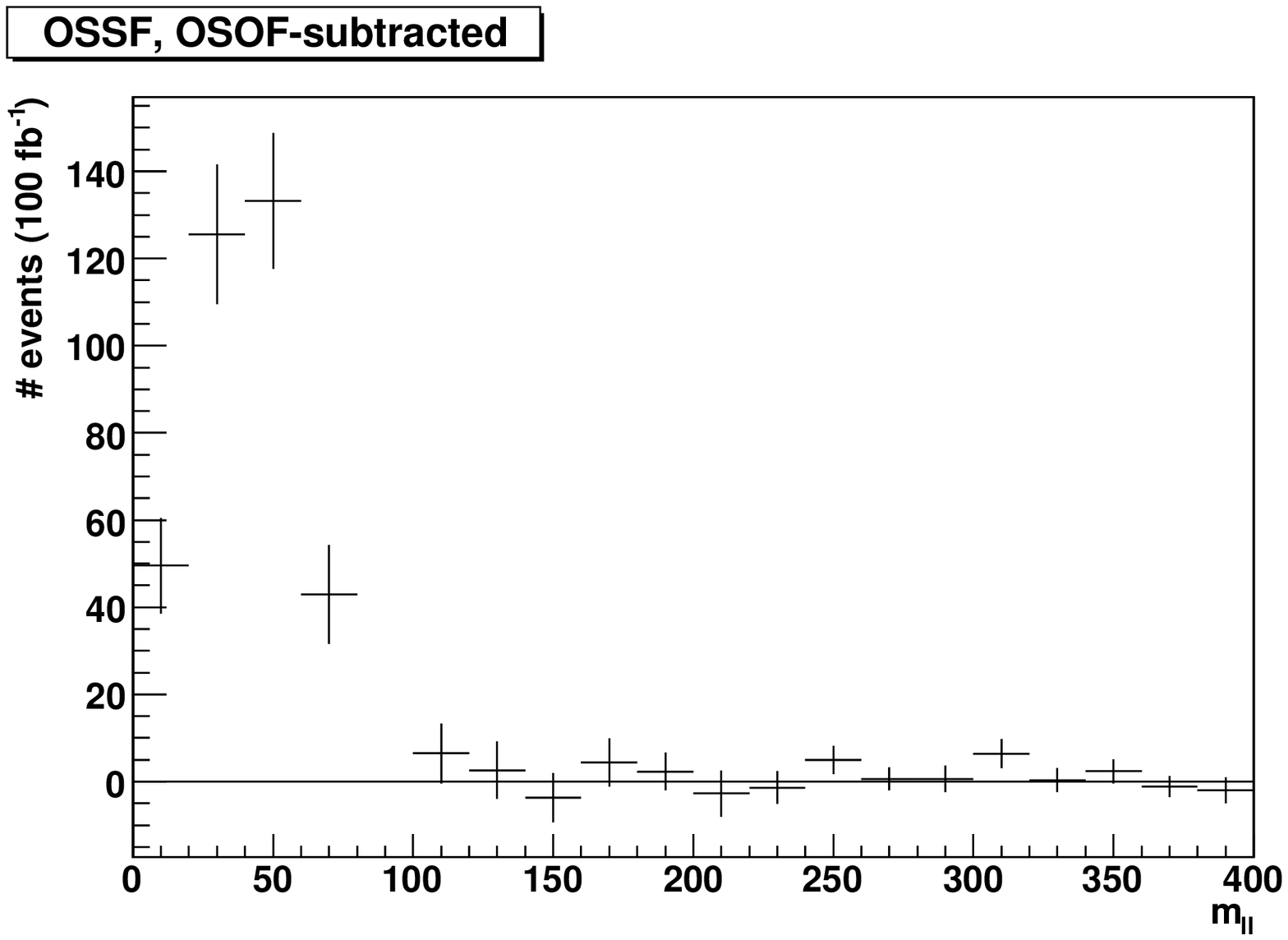}
\epsfxsize=0.49\textwidth\epsfbox{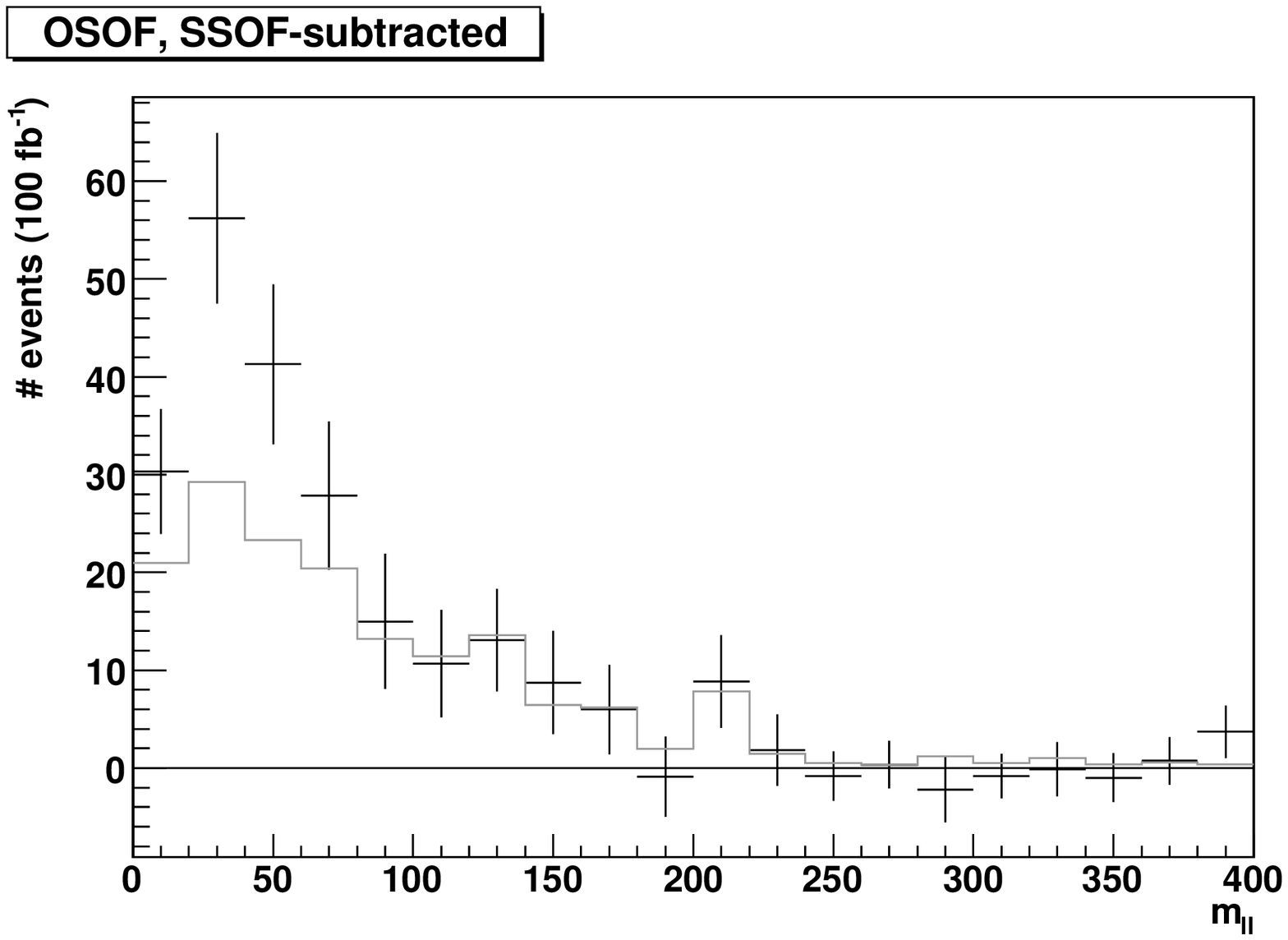}
\caption{(Simple RH-active spectrum.) Dilepton invariant mass distributions applying the 
OSSF-OSOF subtraction (left) and OSOF-SSOF subtraction (right).  Backgrounds are included 
in the subtractions.  The continuous gray histogram is background-only.}
\label{fig:subtractions1}
\end{center}
\end{figure}


\begin{figure}[t]
\begin{center}
\epsfxsize=0.49\textwidth\epsfbox{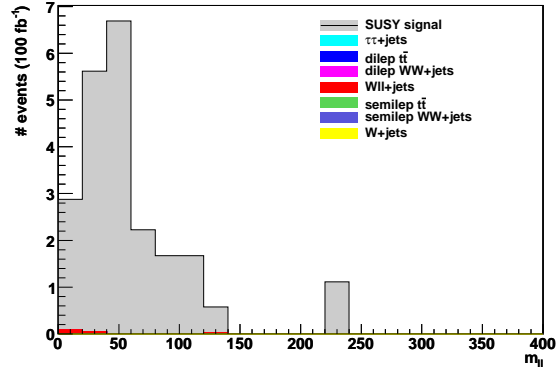}
\caption{(Simple RH-active spectrum.) Dilepton invariant mass distribution within the OFOF trilepton channel. 
We use the procedure described in~\cite{Katz:2009qx} to reduce combinatorial ambiguities.}
\label{fig:mllofof-simple}
\end{center}
\end{figure}


\begin{figure}[t]
\begin{center}
\epsfxsize=0.49\textwidth\epsfbox{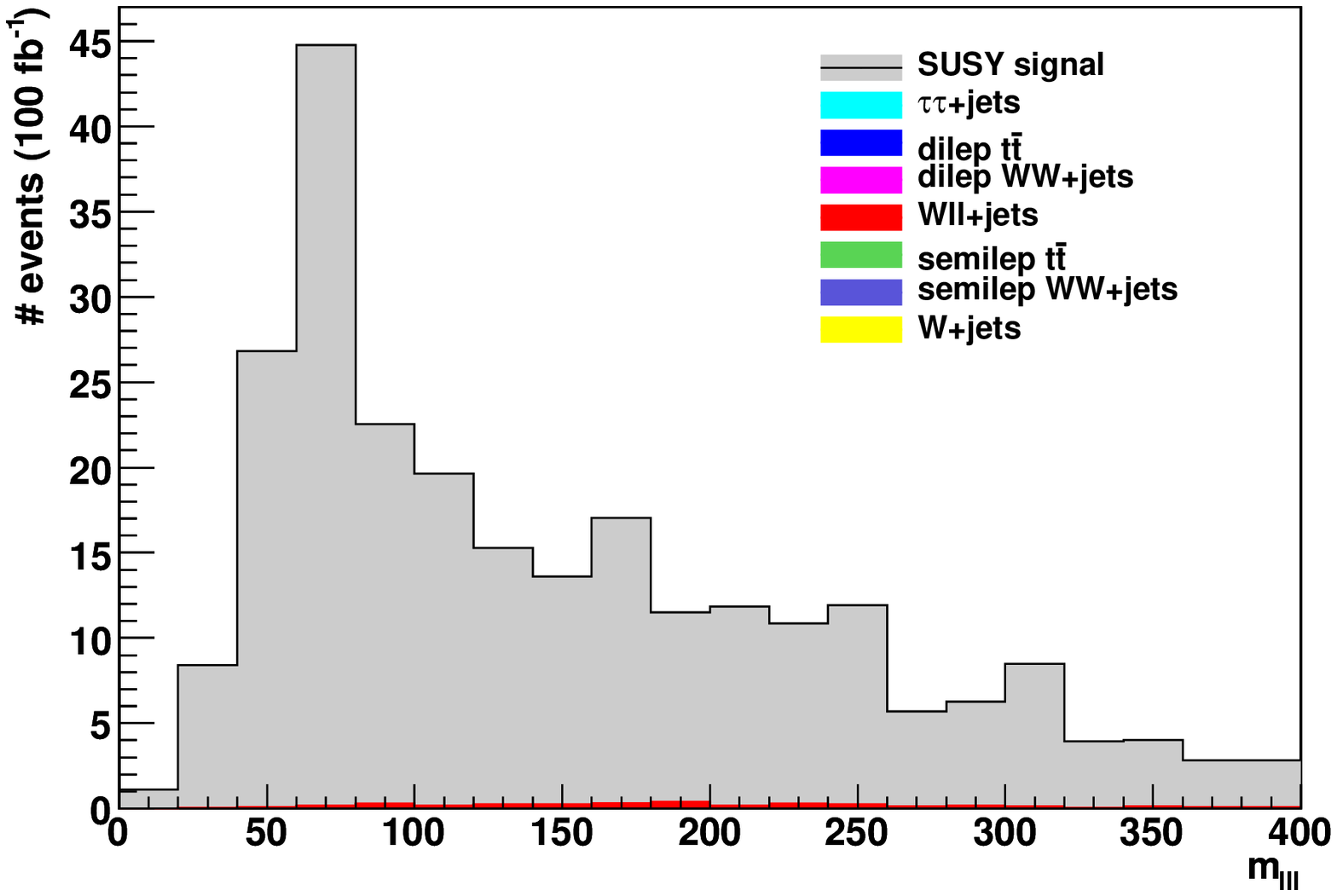}
\epsfxsize=0.49\textwidth\epsfbox{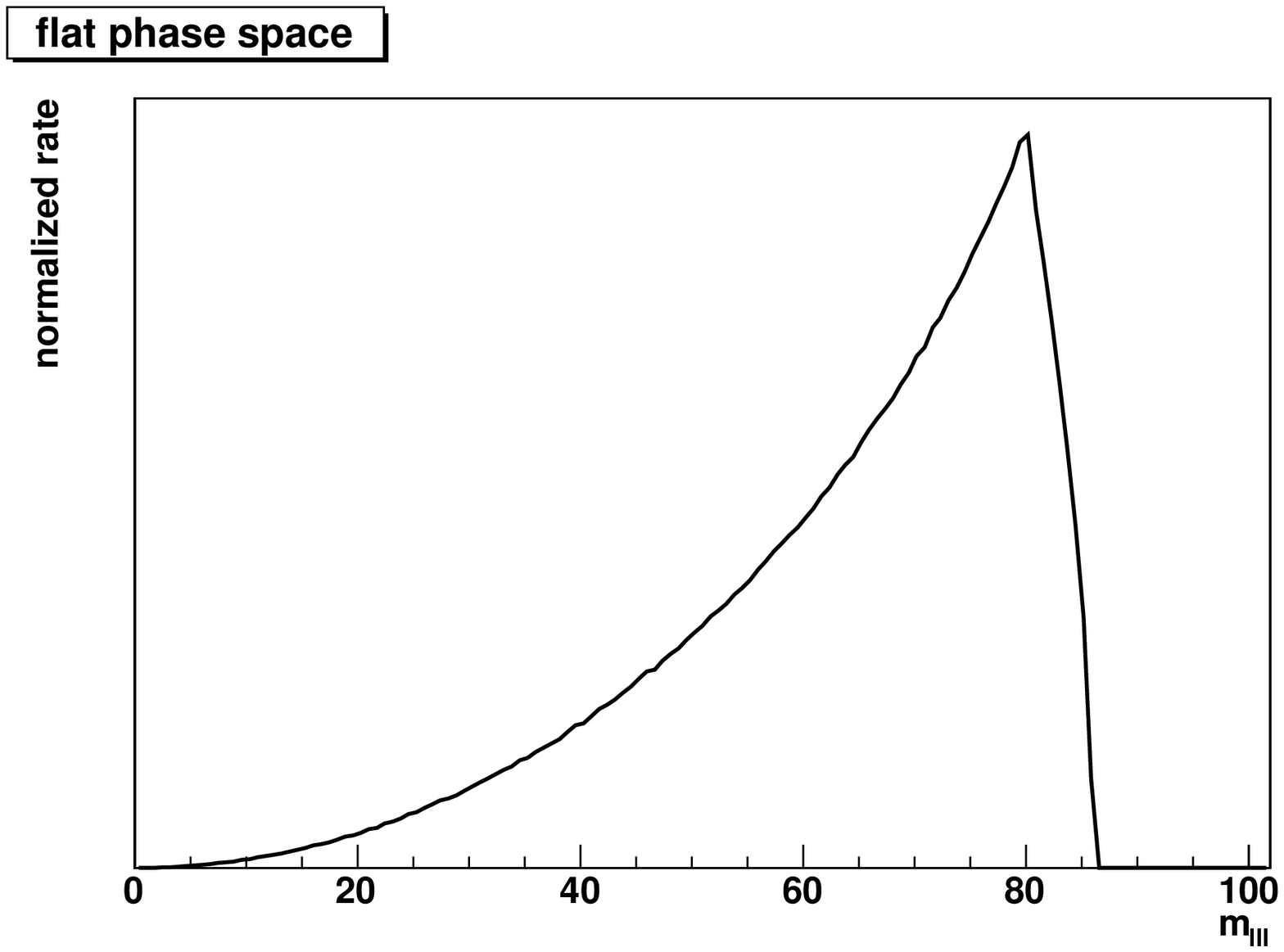}
\caption{(Simple RH-active spectrum.) Trileptonic invariant mass distribution of SFOF and SFSF leptons.
On the right we show a theoretical prediction assuming that all three
leptons are coming from the same decay chain, using a flat phase space generator.  
Note the different horizontal scales.}
\label{fig:mlll-simple}
\end{center}
\end{figure}

We start by analyzing a simple RH-active spectrum, where left-right mixing effects
in the first two generations can be largely neglected, and electron-muon asymmetries
are small.  This spectrum was simulated
using the assumptions of General Gauge Mediation~\cite{Meade:2008wd}, with a 
mediation scale of 100~TeV and $\tan\beta = 5$.  The first column of table~\ref{tab:output_invlep} displays
the physical mass spectrum. 
Since the Bino-like neutralino is relatively light (285 GeV) and there is no significant 
mass gap between LH and RH sleptons, RH smuons predominantly decay through three-body 
neutralino-mediated processes. 

The total leading-order SUSY production cross-section is 558~fb.  
Of this, 264~fb 
is super-QCD pair production, dominantly $\squark\squark^*$ and $\squark\squark$.

We first show the number of observed leptons (Fig.~\ref{fig:nl-simple}). As expected, 
the spectrum exhibits rich leptophilic behavior, with roughly 100 4-lepton events
and handful of 5- and 6-lepton events.

The structure of the dileptonic and trileptonic channels is already very 
suggestive at the level of simple counting (Fig~\ref{fig:structure1}). The dileptonic 
signal appears with OSSF$>$OSOF$>$SSOF$\simeq$SSSF.
The trileptonic channel includes a significant amount of OFOF events, which indicates
the presence of OS, flavor-uncorrelated lepton pairs.  Moreover,
summing the contents of OFOF and SFSF approximately matches the content of SFOF.
This agrees with our expectations for RH-active spectra, as discussed in 
subsection~\ref{3lsignals}.

We next analyze the dilepton invariant mass distributions using subtractions.
We expect to see kinematic features from the following subcascades:
\begin{itemize}
\item $\bino   (\gaugino_1^0) \to \sel  \to \snu$: a bump in the OSSF-OSOF subtraction with endpoint at 89~GeV (peak near 45~GeV),
\item $\bino   (\gaugino_1^0) \to \slep \to \snu$: a bump in the OSOF-SSOF subtraction peaked at 40~GeV,
\item $\wino^0 (\gaugino_2^0) \to \slep \to \snu$: a bump in the OSOF-SSOF subtraction peaked at 60~GeV.
\end{itemize}

The dilepton invariant mass distributions in the different 2l channels (Fig.~\ref{fig:mll-simple})
indeed confirm that OSSF and OSOF dileptons have different shapes.  Both are significantly
above the SM backgrounds.  Unfortunately, the expectation concerning the OSSF endpoint cannot be 
cleanly verified in this case since it falls in the bin surrounding the $Z$-mass, which has been 
intentionally blinded to avoid contamination from the possible SUSY production of on-shell $Z$s. 
Nevertheless the shape of the distribution is clearly visible after subtraction 
(Fig.~\ref{fig:subtractions1}).  Regarding the OSOF distribution, which is
crucial evidence for sneutrino NLSP, we can clearly identify the first peak after subtraction
(Fig.~\ref{fig:subtractions1}).  The second 
peak, however, cannot be clearly distinguished. 
The first peak also materializes in the trileptonic OFOF channel (Fig.~\ref{fig:mllofof-simple}),
albeit with rather low statistics.

Finally, we look at the trilepton invariant mass of SFOF and SFSF dileptons (Fig.~\ref{fig:mlll-simple}).
If all three leptons were always coming from the same decay chain, corresponding to the process
$\bino\to \sel \to \bino^* \to \slep$, they would reproduce, up to spin effects, 
the distribution depicted on the right side of
Fig.~\ref{fig:mlll-simple}.  Although realistically we have combinatorial backgrounds, as well as
backgrounds from LH chains (which we have not subtracted, due to limited statistics),  we can 
still see clear evidence
for the sharp spike.  Observation of such a feature should motivate a more detailed kinematic
analysis, which in this case is quite nontrivial.  (However, see~\cite{Miller:2005zp} for formulas applying
to the case where the intermediate neutralino in $\sel$ decay is on-shell.)

\subsection{RH-Active Spectrum with Flavor Non-Universal Signals due to Smuon Left-Right Mixing}

We next analyze a spectrum where left-right mixing of the smuon leads to appreciably flavor 
non-universal signals.  The physical spectrum appears in the second column of 
table~\ref{tab:output_invlep}.  The mass gap between
the RH and LH sleptons is now large enough to allow for real emission of electroweak
gauge bosons and Higgs bosons.  This spectrum was again produced using General
Gauge Mediation at a mediation scale of 100~TeV, and with $\tan\beta=20$.  
The total leading-order SUSY cross section 
is 911~fb, of which 524~fb is from super-QCD pair production.  SQCD production 
is again squark-dominated.

One additional noteworthy feature of this spectrum is that the neutralino mixing
is not small, and the Bino is distributed nontrivially between the first and third
neutralinos.  Specifically, 
the gauge eigenstate composition of the lightest neutralino (360~GeV) is 
(36\% $\bino$, 4\% $\wino$, 33\% $\higgsino_d$, 27\% $\higgsino_u$),
and of the third neutralino 
(64\% $\bino$, 5\% $\wino$, 15\% $\higgsino_d$, 17\% $\higgsino_u$).\footnote{The 
complete neutralino mixing matrix, with ($\bino$,$\wino$,$\higgsino_d$,$\higgsino_u$) 
composition running horizontally and mass states increasing as we move down vertically, is 
$\left[ \begin{array}{rrrr}  0.60 & -0.20 &  0.57 & -0.52 \\ 
                            -0.04 &  0.05 &  0.70 &  0.71 \\ 
                             0.80 &  0.22 & -0.39 &  0.41 \\ 
                            -0.06 &  0.95 &  0.17 & -0.24  \end{array}  \right]$.   }  
This substantive mixing ultimately ends up having no qualitative impact
on our analysis.  Though there are now naively two Binos in the cascades,
with RH squarks (of the first two generations) 
decaying 40/60\% of the time into the first/third state, the
subsequent decays of the lightest neutralino are highly biased towards invisible
$\nu\snu$ modes due to Wino-Bino interference combined with small phase space for
$l\sel$.  Effectively, then, there is only one Bino ($\gaugino^0_3$), and it is
produced with a somewhat attenuated rate.  This kind of situation is of course
not required to achieve large electron-muon asymmetries, but it illustrates that
our signals can actually be quite robust against neutralino mixing, despite the
fact that most of our discussions above (and in~\cite{Katz:2009qx}) used a simplified 
picture with pure gauge eigenstates.

\begin{figure}[t]
\begin{center}
\epsfxsize=0.49\textwidth\epsfbox{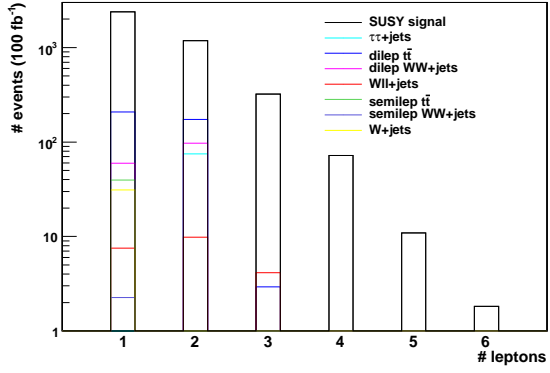}
\caption{(LR-mixed smuon spectrum.) Multilepton counting, using
a logarithmic scale.}
\label{fig:nl-gap}
\end{center}
\end{figure}

\begin{figure}[t]
\begin{center}
\epsfxsize=0.49\textwidth\epsfbox{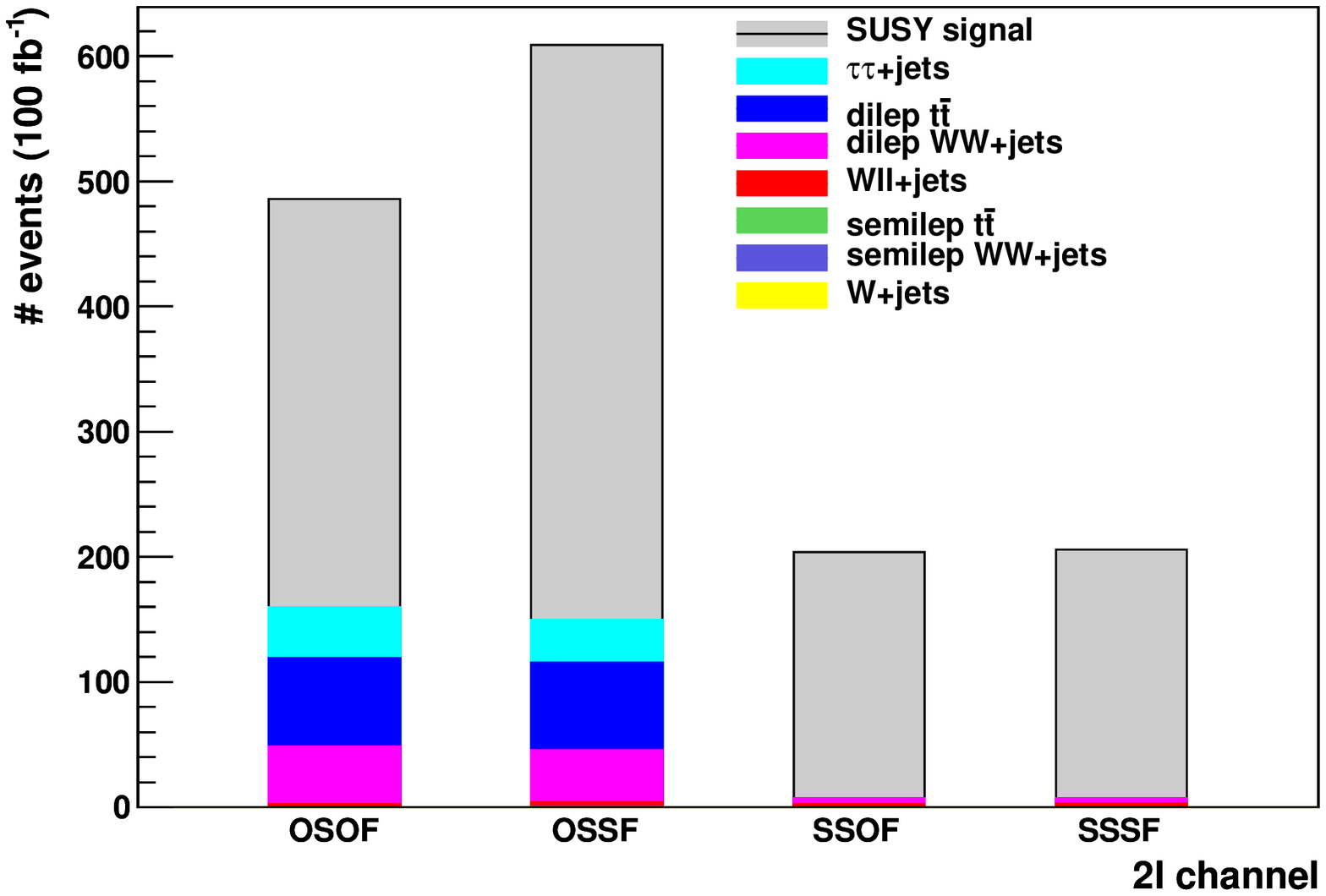}
\epsfxsize=0.49\textwidth\epsfbox{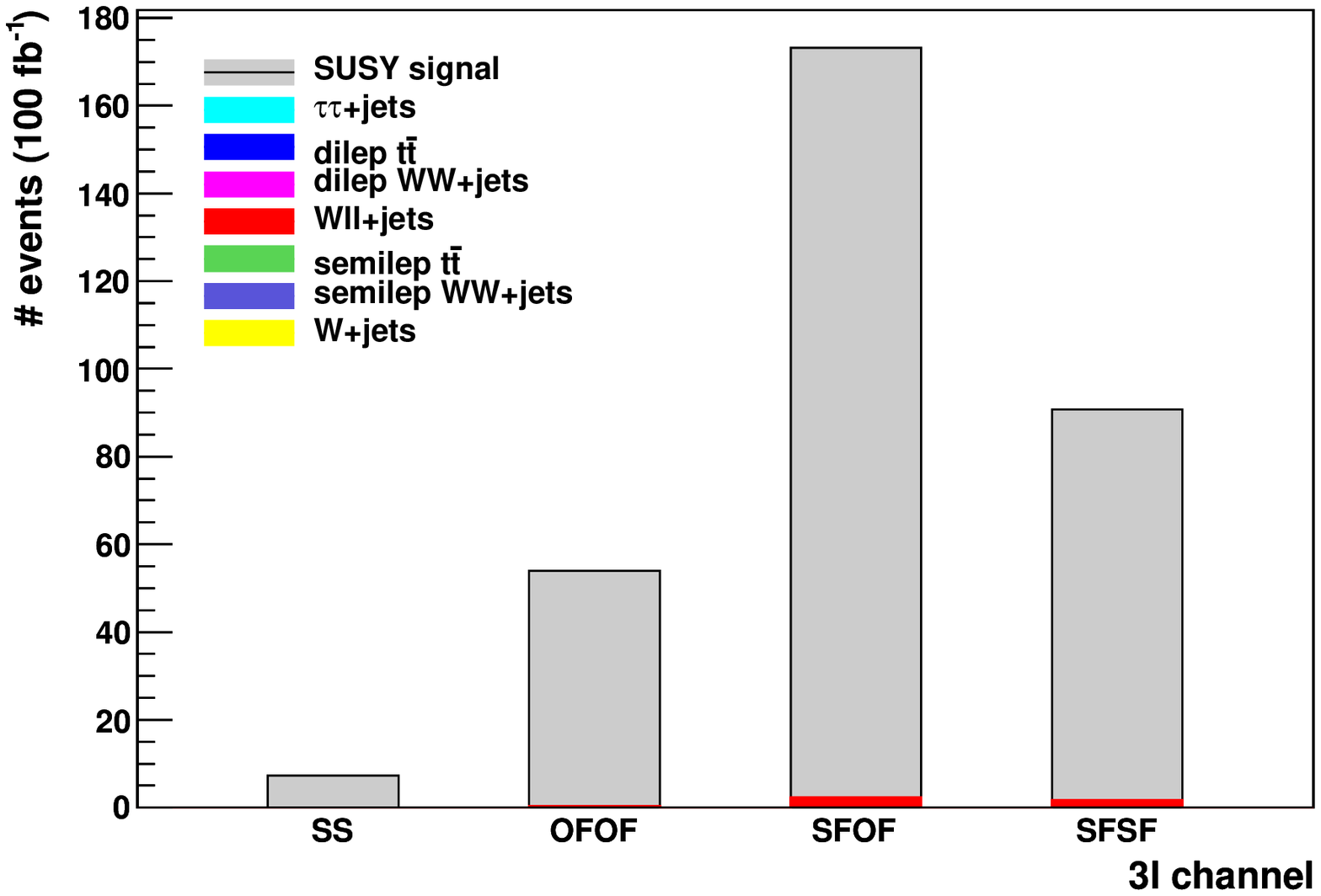}
\caption{(LR-mixed smuon spectrum.) Relative sign/flavor structure of the dileptonic and 
trileptonic channels.  Histograms are stacked. }
\label{fig:structure_gap}
\end{center}
\end{figure}

\begin{figure}[t]
\begin{center}
\epsfxsize=0.49\textwidth\epsfbox{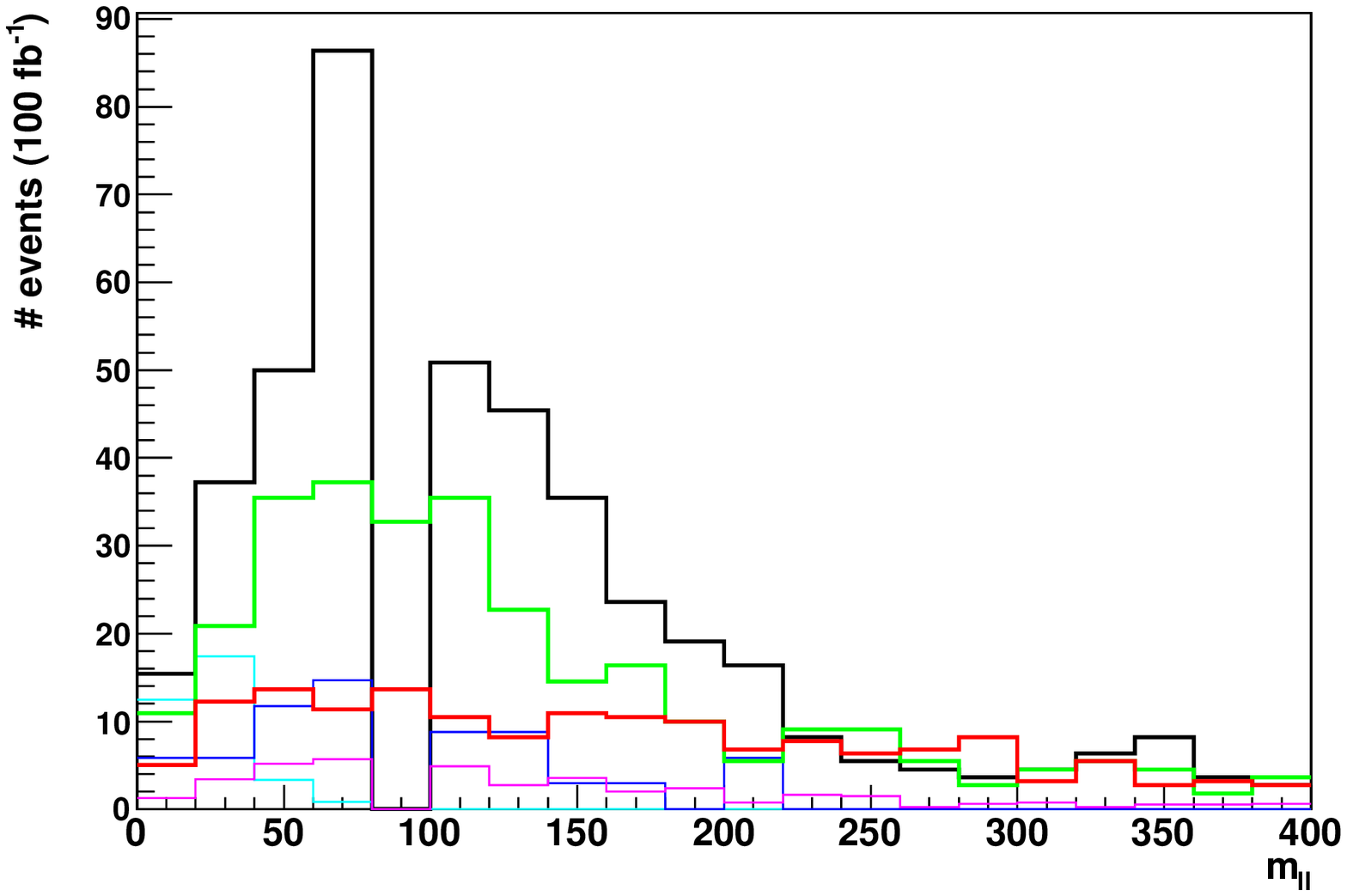}
\caption{(LR-mixed smuon spectrum.) Dilepton invariant mass distributions.
The signal histograms are OSSF (black), OSOF (green), and averaged SS (red).
The major backgrounds, averaged between OSSF and OSOF, are $\tau\tau$ (cyan), dileptonic
$t\bar t$ (blue), and dileptonic $WW$ (pink).}
\label{fig:mll-gap}
\end{center}
\end{figure}

\begin{figure}[t]
\begin{center}
\epsfxsize=0.49\textwidth\epsfbox{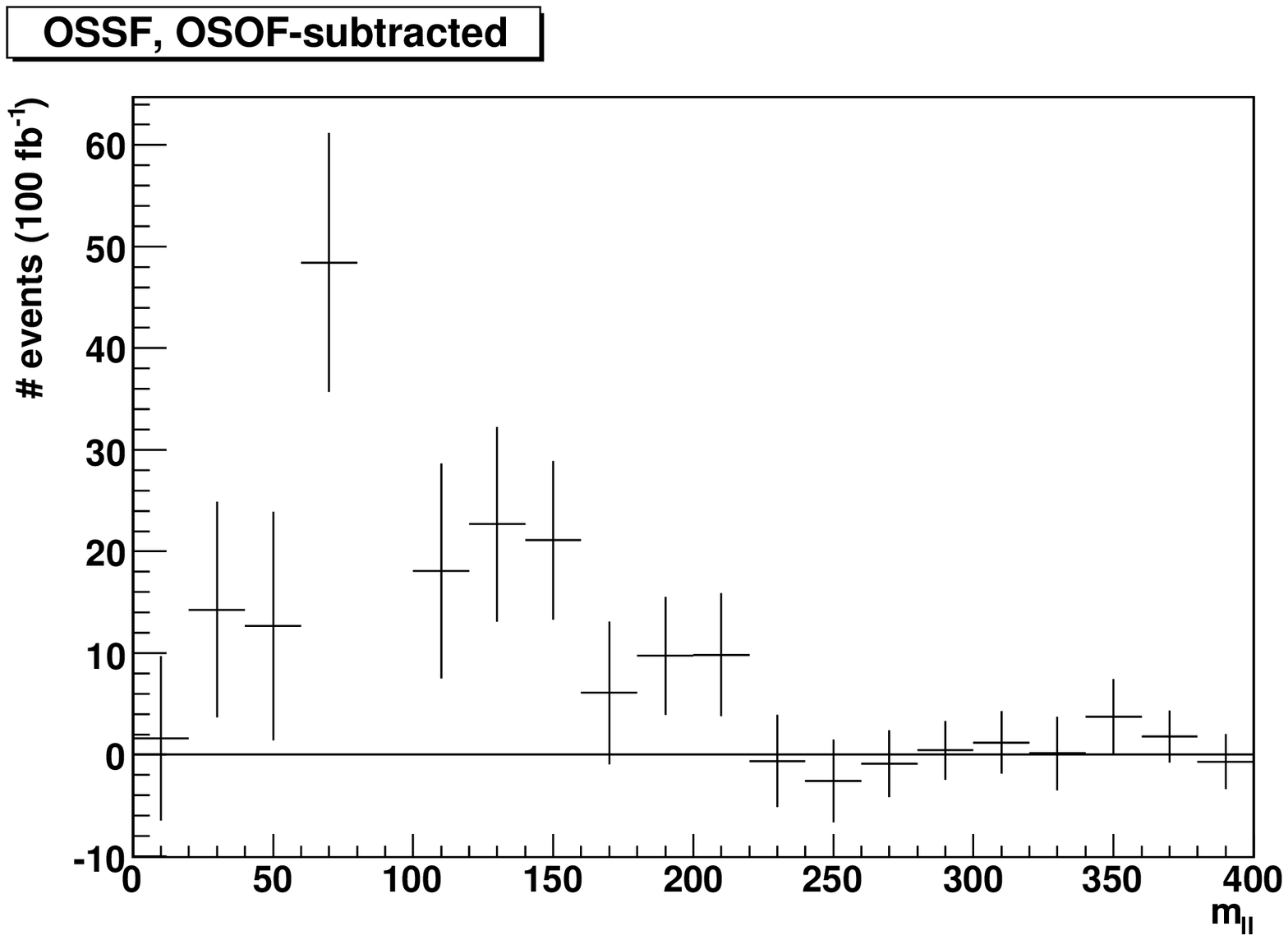}
\epsfxsize=0.49\textwidth\epsfbox{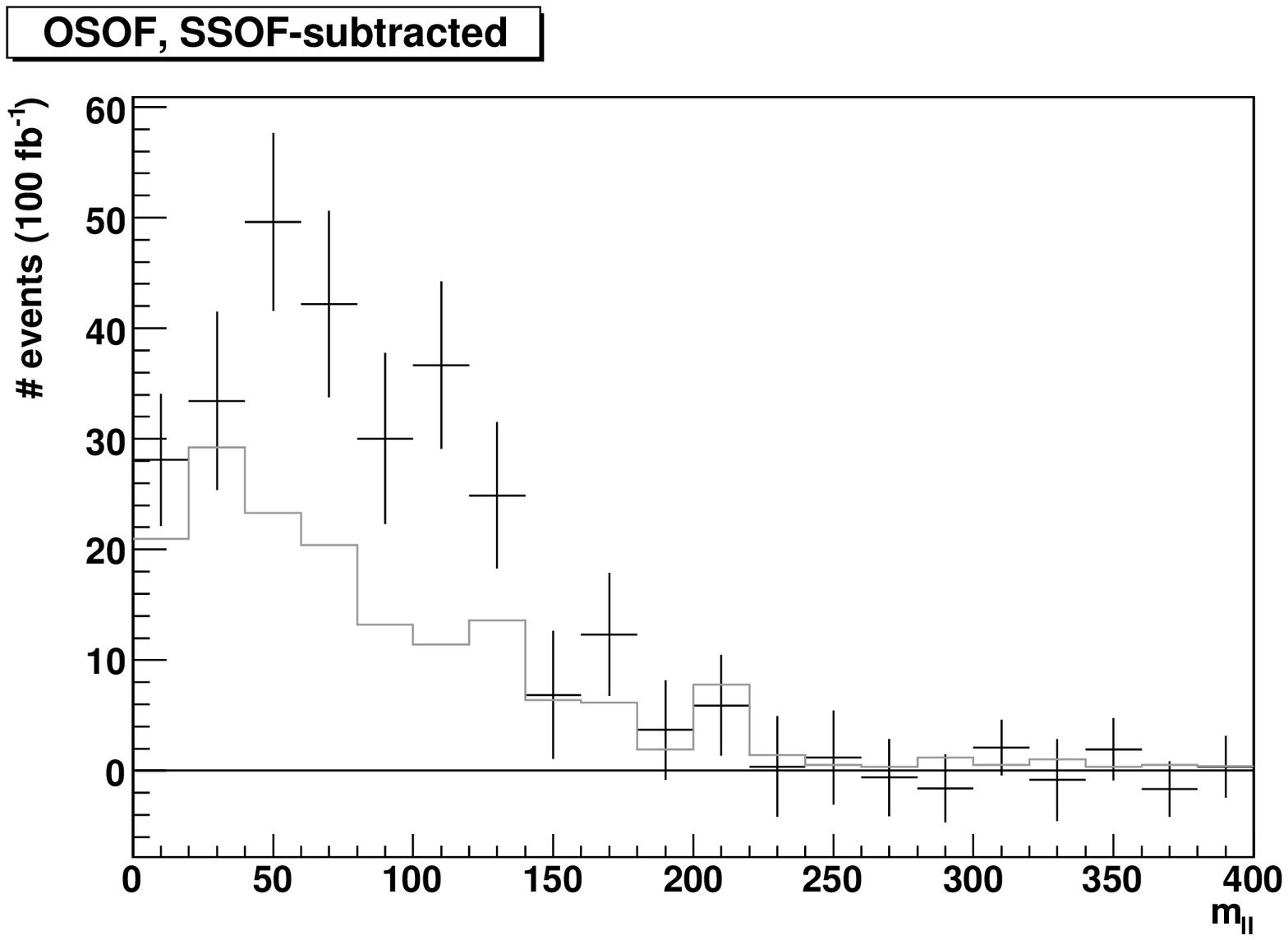}
\caption{(LR-mixed smuon spectrum.)  Dilepton invariant mass distributions applying the OSSF-OSOF subtraction
(left) and OSOF-SSOF subtraction (right).   Backgrounds are included 
in the subtractions.  The continuous gray histogram is background-only.}
\label{fig:subtractions-gap}
\end{center}
\end{figure}

We see from Figs.~\ref{fig:nl-gap} and~\ref{fig:structure_gap} that the 
lepton counting in this spectrum, neglecting detailed electron and muon composition, 
is similar to the simple RH-active spectrum.  In the 
invariant mass distributions, we expect to see following kinematic features:
\begin{itemize}
\item $\bino   (\gaugino_3^0) \to \sel  \to \snu$:  a bump in the OSSF-OSOF subtraction with endpoint at 147~GeV (peak near 75~GeV),
\item $\bino   (\gaugino_3^0) \to \slep \to \snu$:  a bump in the OSOF-SSOF subtraction peaked at  65~GeV,
\item $\wino^0 (\gaugino_4^0) \to \slep \to \snu$:  a bump in the OSOF-SSOF subtraction peaked at 120~GeV.
\end{itemize}
The first two of these predictions are indeed observed (Figs.~\ref{fig:mll-gap} and~\ref{fig:subtractions-gap}),
though statistics are somewhat limited.

\begin{figure}[t]
\begin{center}
\epsfxsize=0.49\textwidth\epsfbox{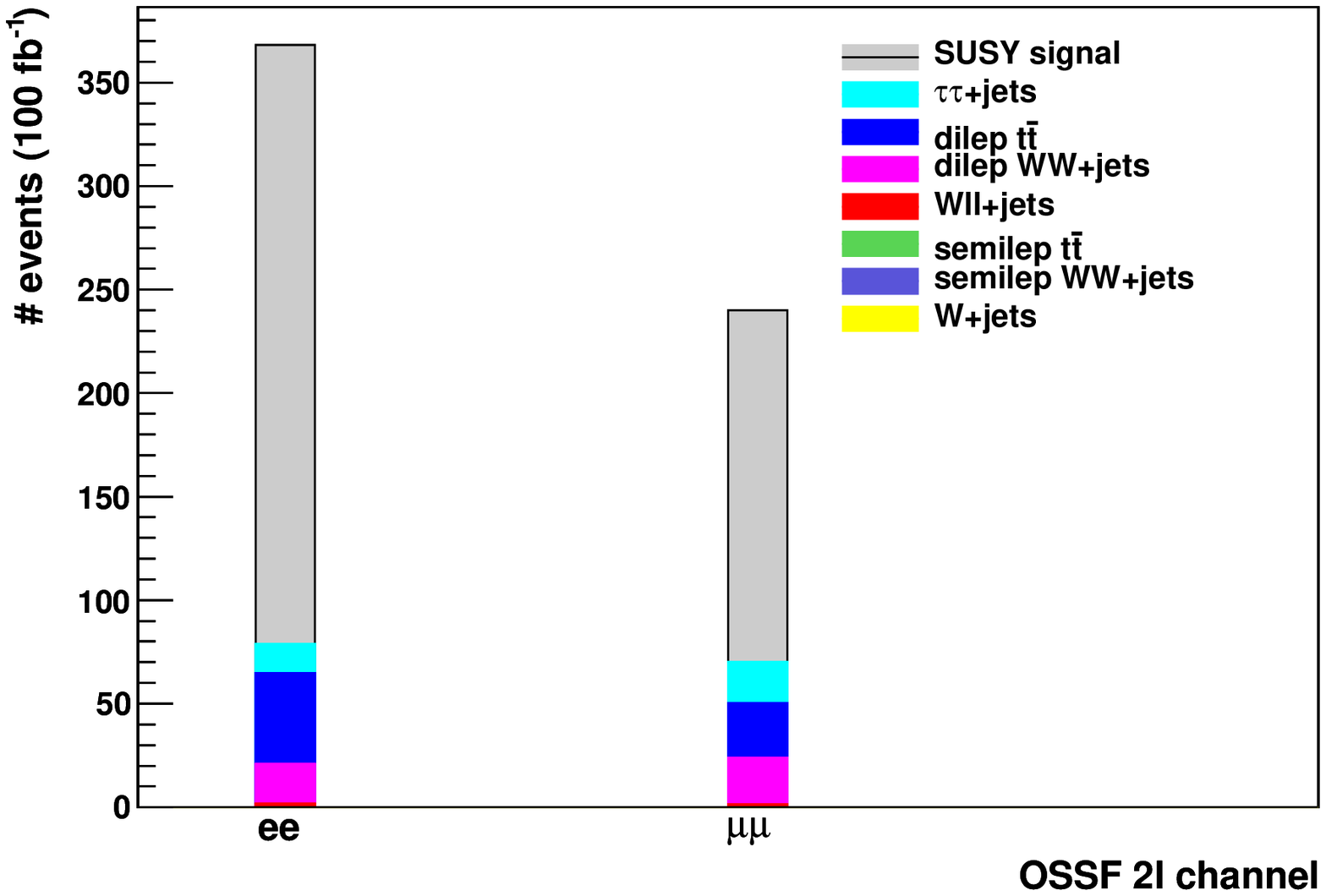}
\epsfxsize=0.49\textwidth\epsfbox{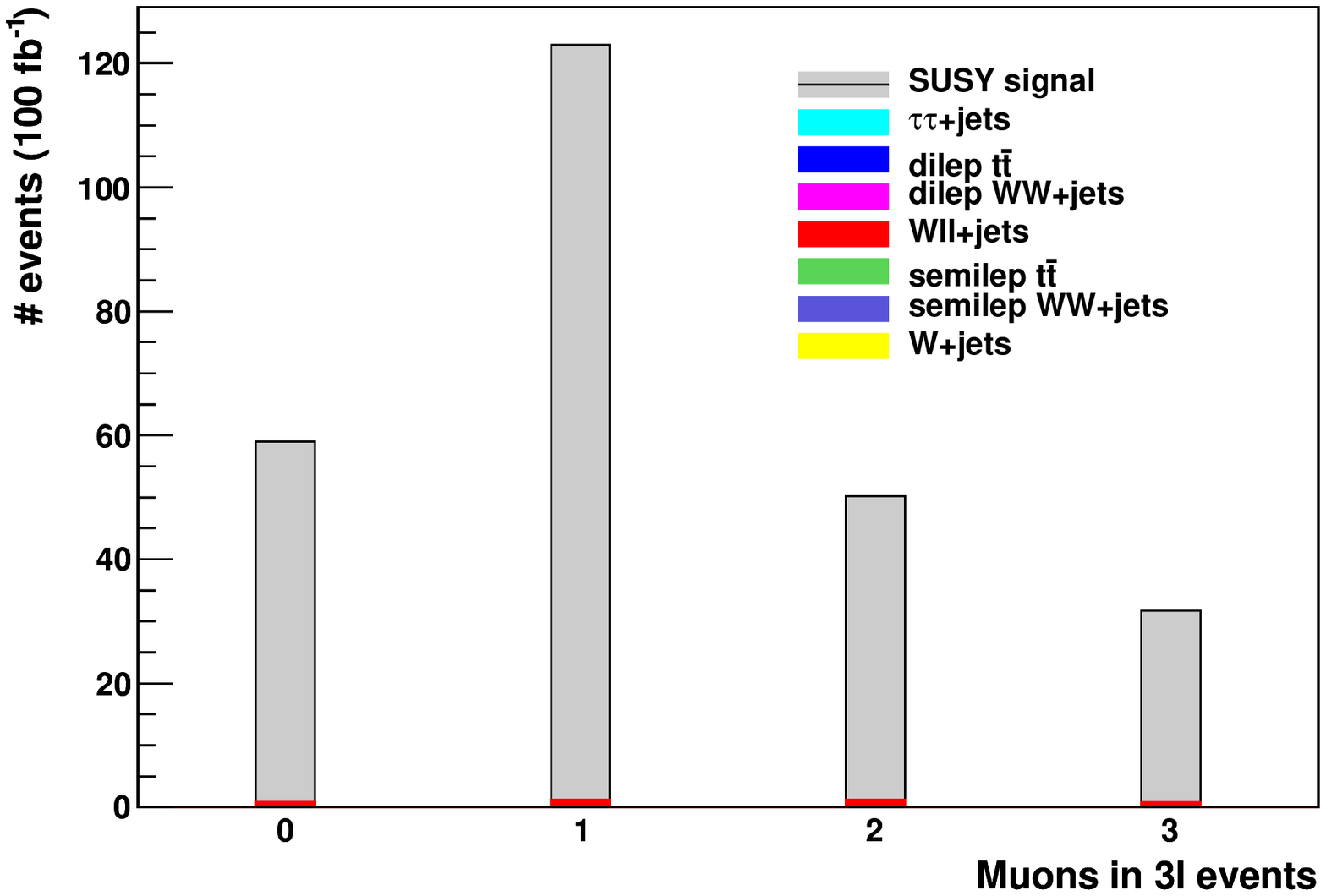}
\caption{(LR-mixed smuon spectrum.)  Electron-muon counting in the dileptonic
OSSF channel (left) and trileptonic SFOF+SFSF channels (right).}
\label{fig:emu_assymetry}
\end{center}
\end{figure}

The most striking new feature of this spectrum is the strong electron-muon
asymmetry, originating from the very different decays of RH selectrons and
RH smuons.  While the former almost exclusively undergoes the usual 3-body
decays mediated by off-shell neutralinos (contributing to the OSSF dilepton 
excess), the latter can mix into a LH
smuon and emit an electroweak gauge boson, or emit a Higgs boson directly
through the Yukawa coupling (subsection~\ref{lrmixing}).  Indeed, we find that 2-body electroweak modes
account for approximately 95\% of the decays of the RH smuon in this spectrum, 
with 63\% going through $W$, and 16\% each through $Z$ and $h$.  We present
simple counting measures of the asymmetry in dileptonic OSSF and trileptonic
SFOF+SFSF in Fig.~\ref{fig:emu_assymetry}.  We can clearly see a large mismatch between
the number of $e^+e^-$ and $\mu^+\mu^-$ events contributing to the total 
(unsubtracted) OSSF sample, without performing any more sophisticated analysis.
In trilepton, we see the asymmetries $0\mu > 3\mu$ and $1\mu > 2\mu$ 
predicted in subsection~\ref{flvor-bias}.

\begin{figure}[t]
\begin{center}
\epsfxsize=0.49\textwidth\epsfbox{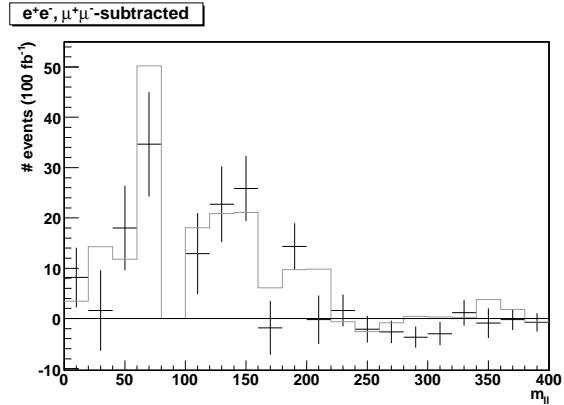}
\caption{(LR-mixed smuon spectrum.)  Dilepton invariant mass distribution applying the
$e^+e^-$-$\mu^+\mu^-$ subtraction.  For comparison, the OSSF-OSOF subtraction is also displayed, as
the continuous gray histogram (without error bars).}
\label{fig:ee_sub}
\end{center}
\end{figure}

We can also plot the difference between the dileptonic invariant mass distributions
in the $e^+e^-$ and $\mu^+\mu^-$ channels (Fig.~\ref{fig:ee_sub}).  While statistical
fluctuations still limit a detailed comparison, the OSSF electron excess over OSSF muons
is clearly consistent in shape and normalization with the total OSSF-OSOF excess.

\subsection{Spectrum with a Light Stau Doublet - NUHM}

Finally, we consider a spectrum with significantly lighter stau doublet,
$m(\snu_\tau) < m(\tilde\tau_L) < m(\snu_{e,\mu})$.  For this analysis, we use the 
second NUHM spectrum
of~\cite{Covi:2007xj}, which arises from gaugino mediation~\cite{Kaplan:1999ac,Chacko:1999mi} 
with a large down-type Higgs mass at the mediation scale of $2\times 10^{16}$~GeV.
The mass spectrum is displayed in the third column of 
table~\ref{tab:output_invlep}.\footnote{The detailed numerical values of the physical
masses differ somewhat from those in~\cite{Covi:2007xj}.  Presumably, this can be
accounted for by the fact that a different version of {\tt SOFTSUSY} 
(namely {\tt 2.0.10}) is used in that paper, or perhaps the input scale
is slightly different.  In any case, these small differences are largely irrelevant
for our analysis.}  The total SUSY leading-order cross section is 4000~fb, with
1995~fb from super-QCD.  The breakdown in terms of exclusive SQCD pairs is
111~fb $\gluino\gluino$, 937~fb $\gluino\squark$, 357~fb $\squark\squark^*$, 
and 590~fb $\squark\squark$.

\begin{figure}[t]
\begin{center}
\epsfxsize=0.49\textwidth\epsfbox{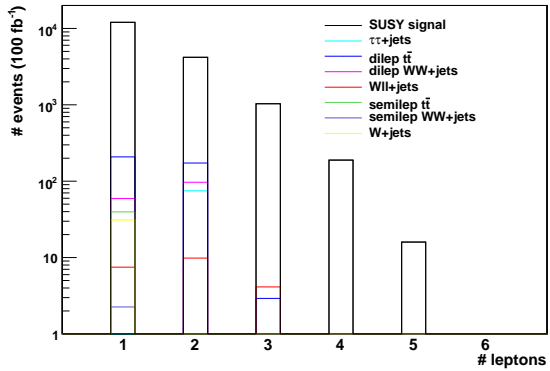}
\caption{(NUHM spectrum.) Multilepton counting, using
a logarithmic scale.}
\label{fig:nl-nuhm}
\end{center}
\end{figure}

\begin{figure}[t]
\begin{center}
\epsfxsize=0.49\textwidth\epsfbox{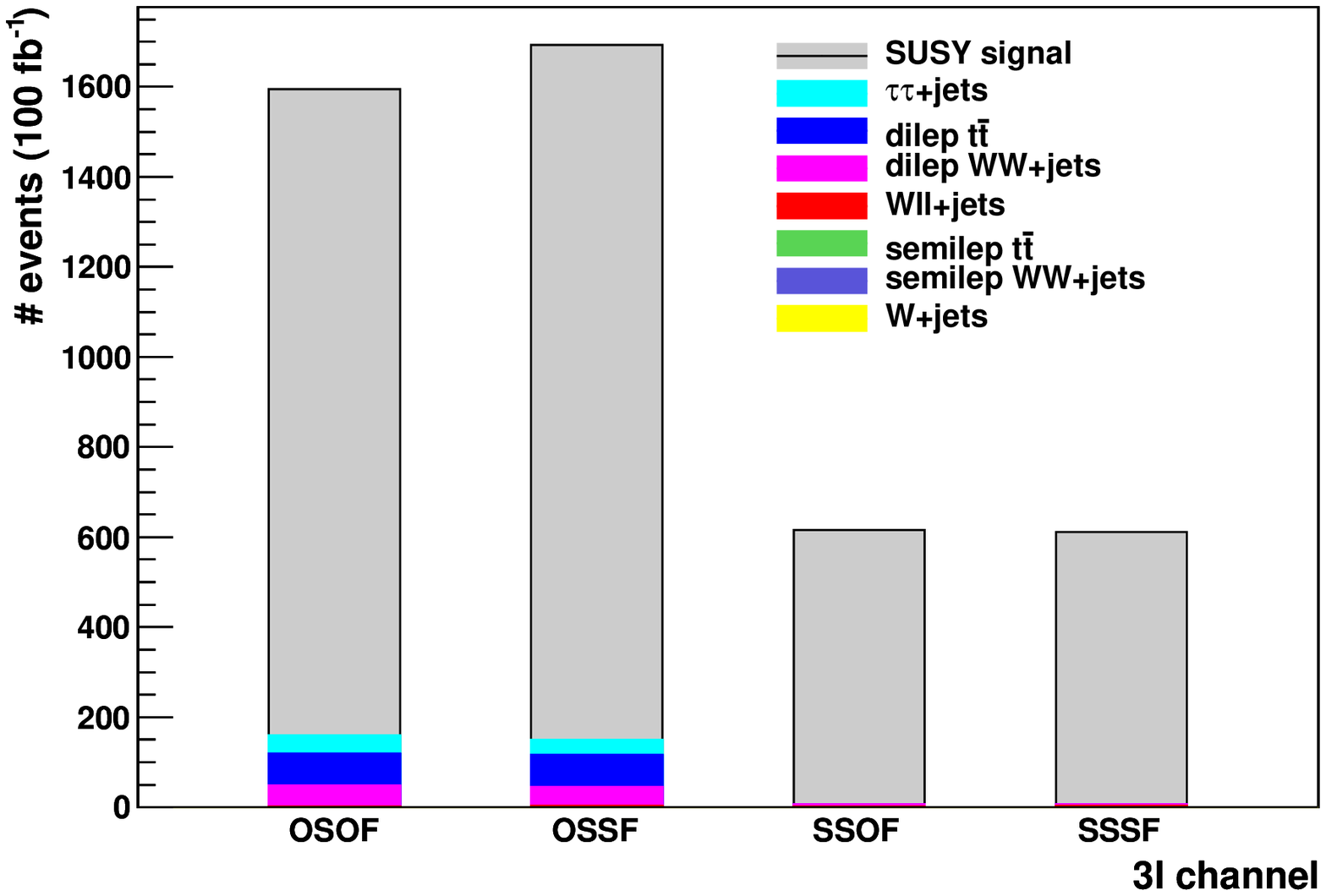}
\epsfxsize=0.49\textwidth\epsfbox{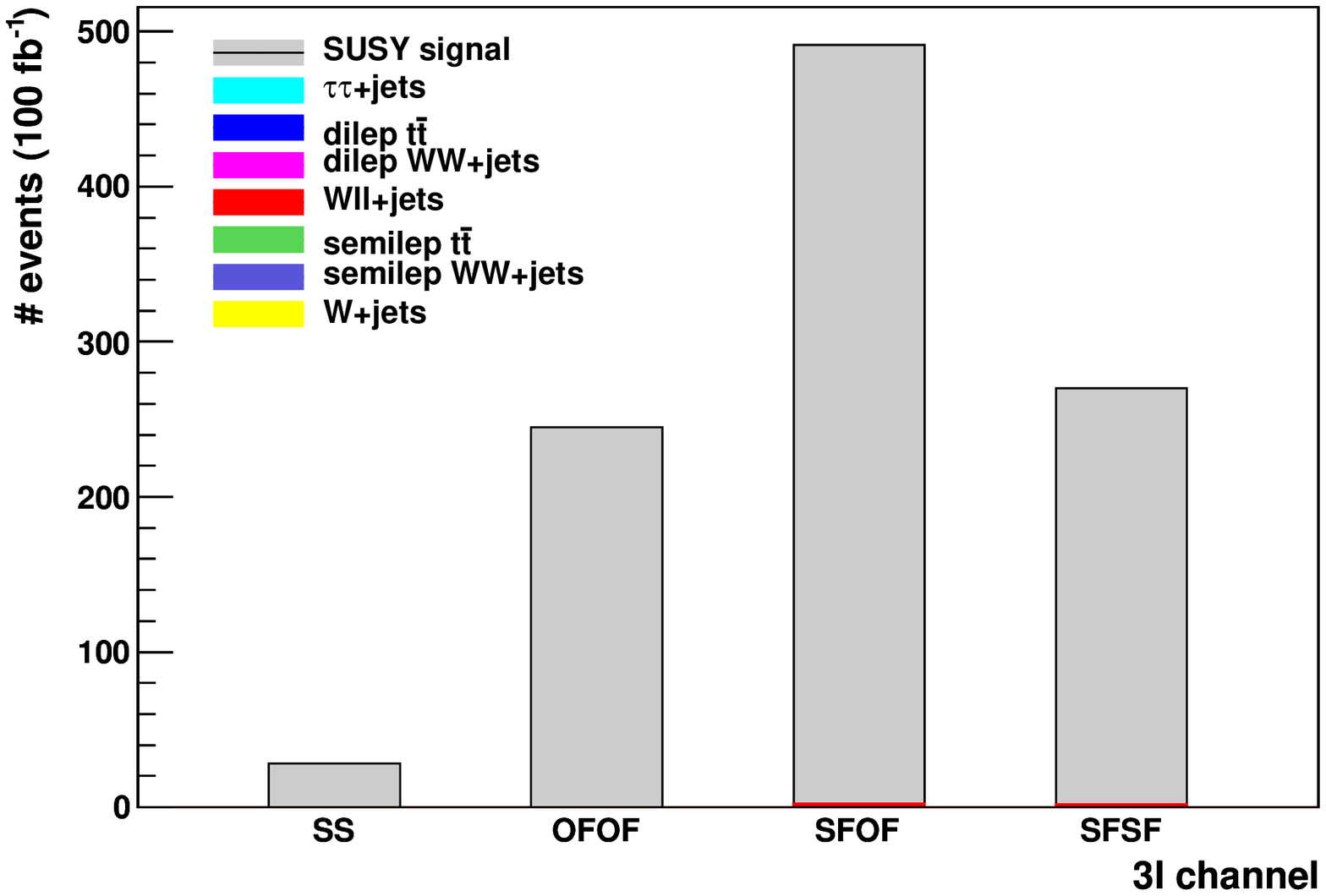}
\caption{(NUHM spectrum.) Relative sign/flavor structure of the dileptonic and 
trileptonic channels.  Histograms are stacked.}
\label{fig:structure_nuhm}
\end{center}
\end{figure}

We can again observe highly leptophilic behavior (Fig.~\ref{fig:nl-nuhm}), though
the presence of OSSF production is now not so obvious from simple counting in
sign/flavor channels (Fig.~\ref{fig:structure_nuhm}).  Indeed, up to some small hint 
of flavor correlation, the latter appears quite consistent with the simplest sneutrino
NLSP spectra.  However, the high multiplicity is already a clear giveaway that there
is more going on.

\begin{figure}[t]
\begin{center}
\epsfxsize=0.49\textwidth\epsfbox{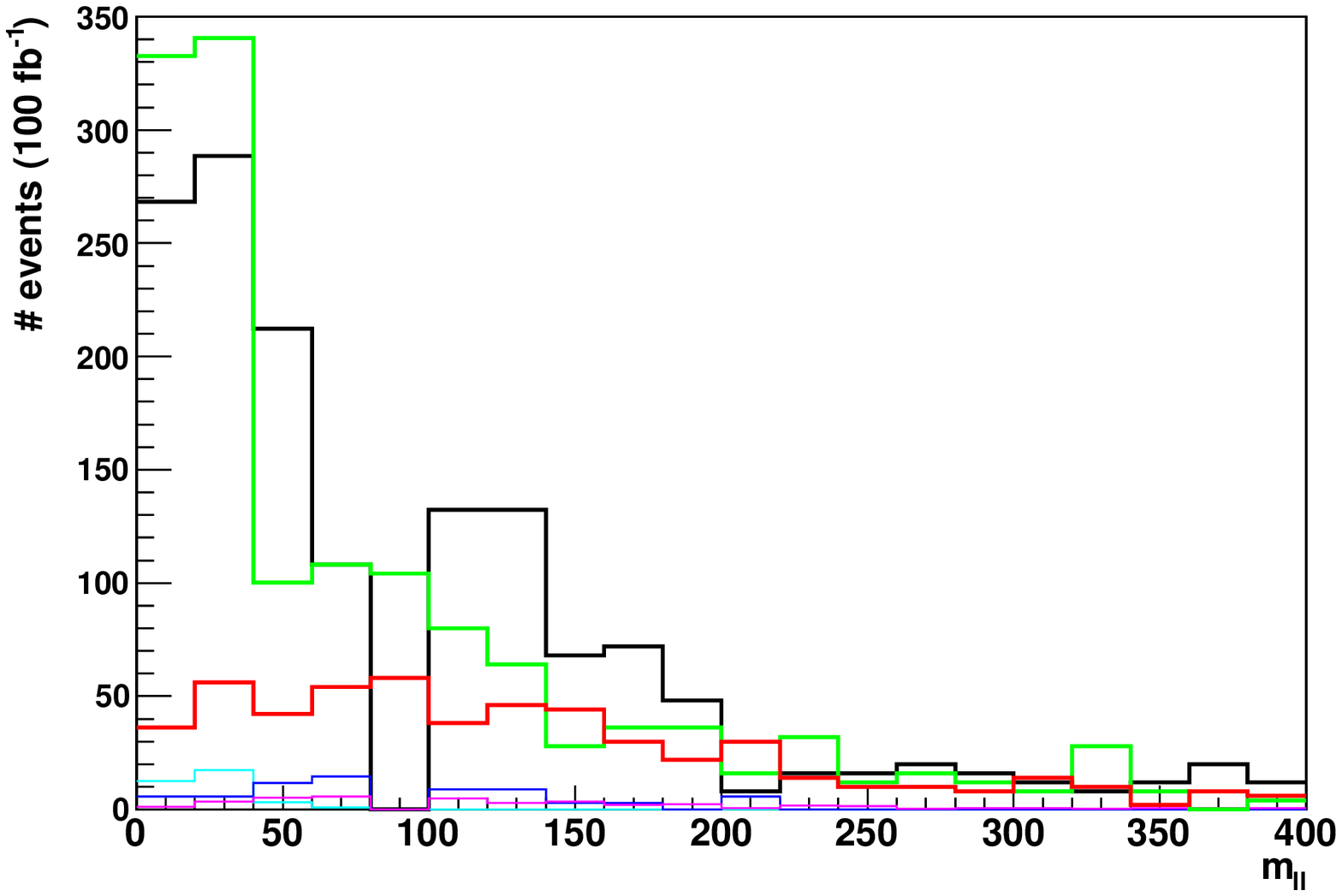}
\caption{(NUHM spectrum.) Dilepton invariant mass distributions.
The signal histograms are OSSF (black), OSOF (green), and averaged SS (red).
The major backgrounds, averaged between OSSF and OSOF, are $\tau\tau$ (cyan), dileptonic
$t\bar t$ (blue), and dileptonic $WW$ (pink).}
\label{fig:mll-nuhm}
\end{center}
\end{figure}

\begin{figure}[t]
\begin{center}
\epsfxsize=0.49\textwidth\epsfbox{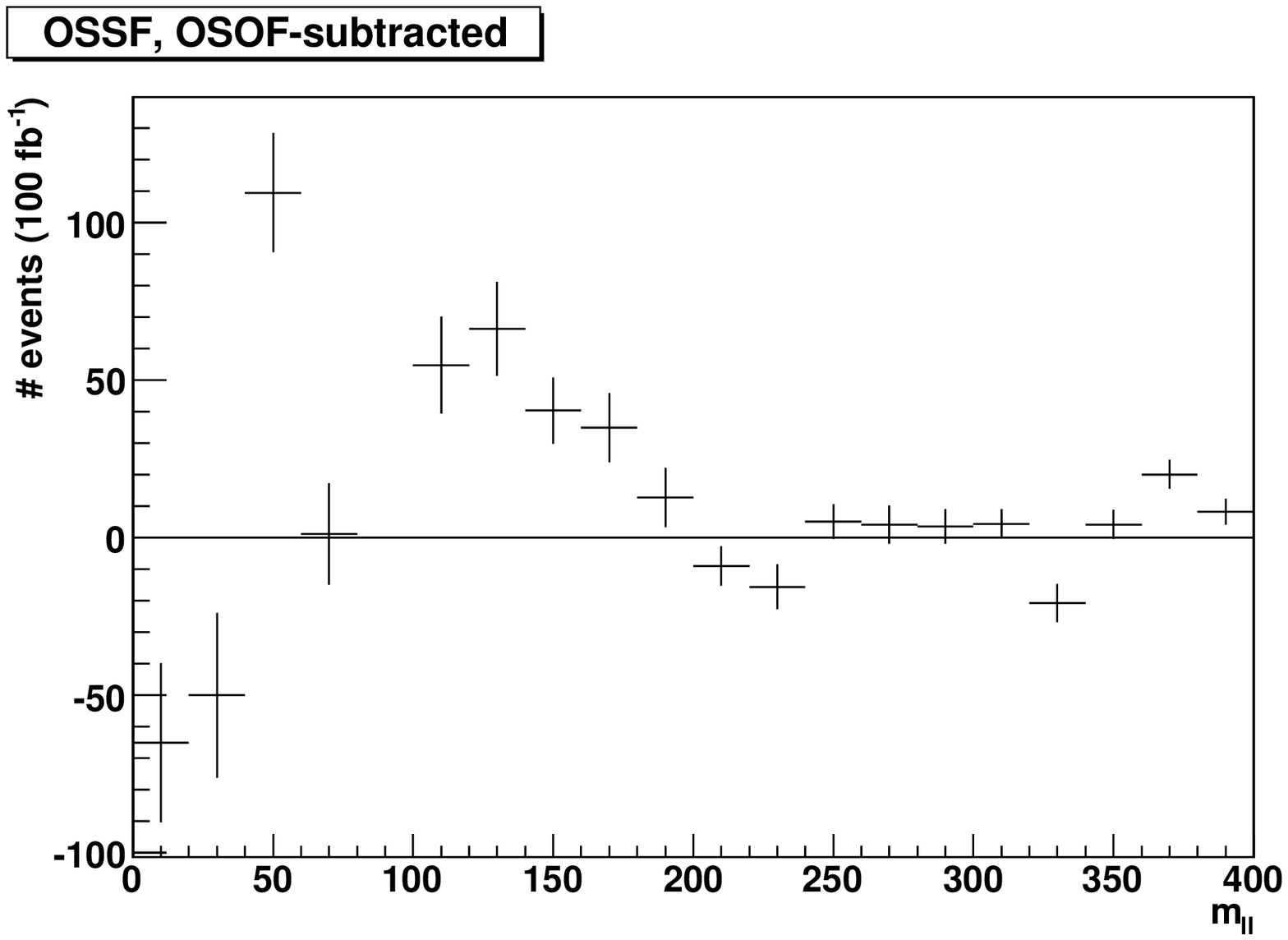}
\epsfxsize=0.49\textwidth\epsfbox{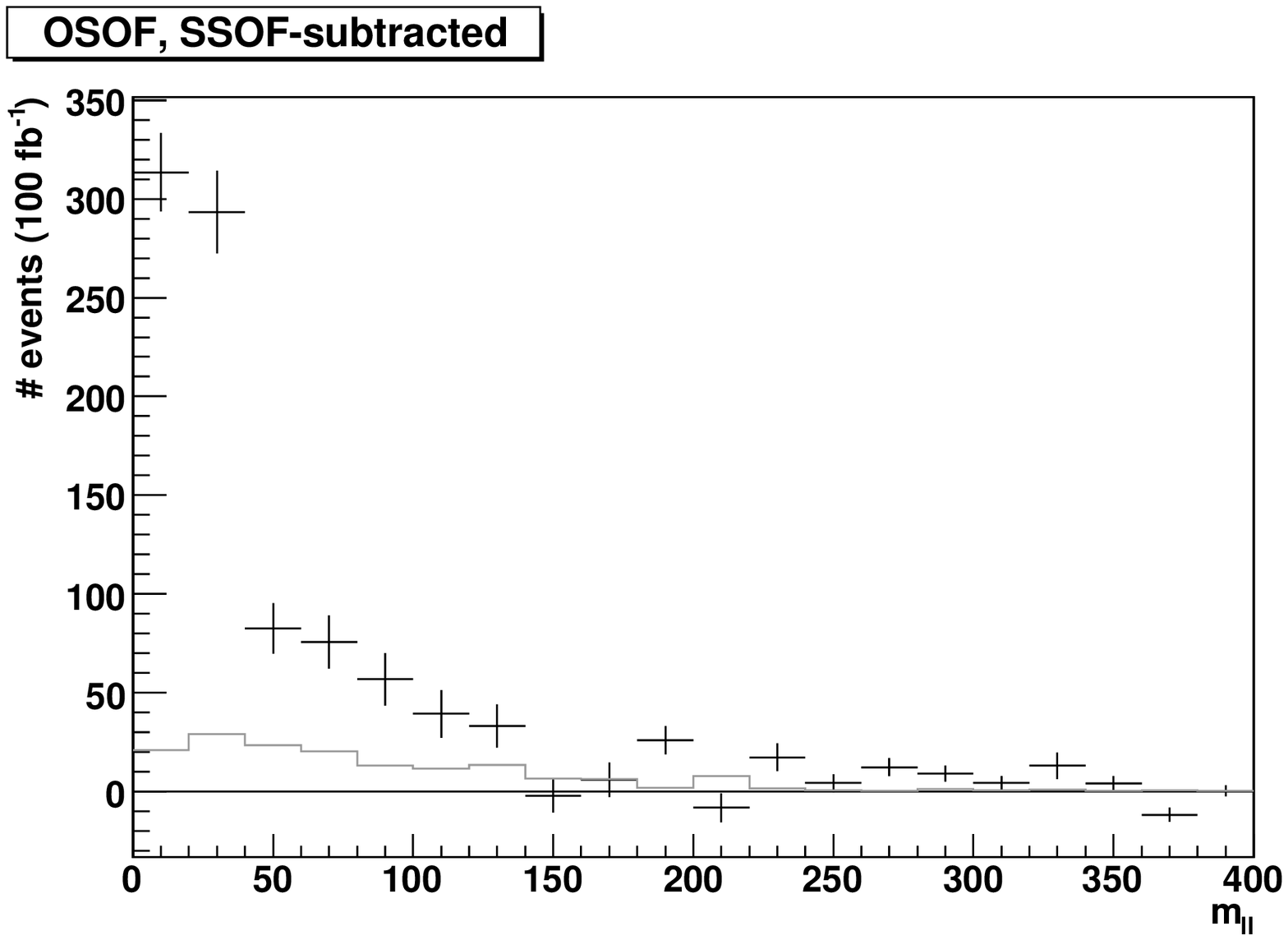}
\caption{(NUHM spectrum.) Dilepton invariant mass distributions applying the OSSF-OSOF subtraction
(left) and OSOF-SSOF subtraction (right).   Backgrounds are included 
in the subtractions.  The continuous gray histogram is background-only.}
\label{fig:subtractions-nuhm}
\end{center}
\end{figure}

Much more can be inferred from the detailed invariant mass distributions.
Given this spectrum, we expect to see following kinematic features in dileptons:
\begin{itemize}
\item $\bino  (\gaugino_1^0) \to \slep \to \snu_\tau   $:  a bump in the OSSF-OSOF subtraction peaked near  77~GeV,
\item $\wino^0(\gaugino_2^0) \to \slep \to \snu_\tau   $:  a bump in the OSSF-OSOF subtraction peaked near 115~GeV,
\item $\bino  (\gaugino_1^0) \to \slep \to \snu_{e,\mu}$:  a bump in the OSOF-SSOF subtraction peaked at 25~GeV,
\item $\wino^0(\gaugino_2^0) \to \slep \to \snu_{e,\mu}$:  a bump in the OSOF-SSOF subtraction peaked at 75~GeV.
\end{itemize}
When we take the distributions (Fig.~\ref{fig:mll-nuhm}) and perform the subtractions 
(Fig.~\ref{fig:subtractions-nuhm}), we see distributions consistent with 
the presence of these
features.  But, as discussed in subsection~\ref{NUHMsignal}, there is also a significant OS
contamination from leptonic tau decays.  This appears in the right panel of 
Fig.~\ref{fig:subtractions-nuhm}
as a high rate in the lowest mass bin, representing an unresolved falling distribution.
Disentangling this from the first slepton/sneutrino bump might be feasible with
higher statistics, and the second bump is already almost well-resolved.  The presence of
such nontrivial structures in the subtracted mass distribution is highly suggestive
of the participation of sleptons and sneutrinos of the first two generations.
However, the evidence is clearly not as clean compared to the cases without significant
tau production.

\section{Conclusions and outlook}\label{conclusions}

In the present work, we have expanded our investigation of spectra with
sneutrino NLSP to incorporate three simple variations which lead to very
rich multilepton plus jets plus missing energy signals
at the LHC.  These include:
\begin{itemize}
\item  spectra with  $m(\sel) < m(\bino) < m(\squark),m(\gluino)$, such
that there is an appreciable
rate for decay chains that produce RH sleptons,  with small 
effects from left-right mixing on the decays of the RH smuon,
\item  spectra of this type where the RH smuon predominantly undergoes
electroweak decays via left-right mixing,
\item  spectra where the LH stau doublet is appreciably lower in mass than
the other LH sleptons.
\end{itemize}

The immediate signals of these spectra are high rates for multileptons, with reasonable chance to
observe up to 6-lepton events in a 100~fb$^{-1}$ run of the LHC.  Beyond this, we can identify
some general features of the dileptonic and trileptonic invariant mass
distributions, which allow
us to extract kinematic information.  Dileptons will contain independent
excesses from the production/decay of
both LH and RH sleptons, the former equally distributed between OSSF and
OSOF channels, and the latter contained
exclusively in OSSF.  Together, these will appear as mismatched OSSF and
OSOF excesses, from which we can
extract the individual distributions by judicious subtractions.
Backgrounds are non-negligible but in principle
manageable.  The same distributions will
also exist in the trilepton channels, along with a potentially observable
excess in the trilepton
invariant mass.  Here, the backgrounds are almost purely combinatorial and
supersymmetric in origin.

In the spectra where electroweak decays dominate for the RH smuon, the
multilepton signals can display
substantial asymmetries between $e$ and $\mu$.  If such signals were
observed at the LHC,
they might initially be
interpreted as large flavor violation in the soft terms of the first two
generations, such
as in~\cite{Kumar:2009sf}.  Amusingly, though, these large effects on the
flavor structure
of the events actually originate directly from the ordinary electron and
muon Yukawa
couplings, appropriately supersymmetrized.

Spectra with a highly split-off stau doublet will look quite similar to
more flavor-degenerate spectra
with RH sleptons participating, though with
a fairly prominent low-mass
OS dilepton feature from leptonically decaying taus.  This would be in
addition to the fact that most
events could be tagged with at least one hadronic tau.

We have seen that these three classes of spectra would be hard to miss at
the LHC, and that they would
support a large number of independent multilepton analyses.
But, if we assume that such signals are supersymmetric in origin, how do
we ultimately know that
they are indicating the presence of a sneutrino NLSP?  The irony of these
scenarios is that
while the high leptonic rates make them easy to discover, they also tend
to obscure the
detailed spectral structure in a pileup of overlapping signals from
different sub-cascades.
Still, the situation remains simple enough in the dileptonic and
trileptonic channels to largely
infer what is happening.

In our first paper, we advocated independent identification
of a flavor-blind OS excess in the dileptonic invariant mass distributions
within both the dilepton and
trilepton channels.  The trilepton analysis provided a powerful
cross-check of the much more
background-contaminated dileptonic analysis.  We can still apply the same
methodology here, but we
must restrict ourselves to the OSOF channel since the OSSF contribution
appears in superposition with
the signal from the RH cascades.  While it will no longer be
straightforward to directly see that the OSOF
signal is indeed half of a flavor-blind OS signal, this remains a very
reasonable first assumption.
In any case, as we have argued in~\cite{Katz:2009qx}, it does not seem straightforward to engineer
such a large OSOF excess within the MSSM.  Naively the easiest way to do
this is to introduce a large population of OSSF taus from stau production and
decay, and these will sometimes
both decay into (flavor-uncorrelated) leptons.  But this kind of scenario
could be inferred in two ways:
the invariant mass distribution would tend to be skewed towards lower
masses due to the energy sharing
with neutrinos, and there would necessarily be a high rate of hadronic tau
production, which could be independently observed.
We would therefore propose that the
observation of the OSOF-SSOF signal with a well-localized invariant mass
bump (or bumps), in combination with all of the other
multileptonic activity, and with a
modest rate for hadronic tau detection, makes the sneutrino NLSP
interpretation highly attractive.

Of course, arguing this case is more subtle for the NUHM-type spectra with
light stau doublet, where
genuine OS dileptons coexist with high production rates for taus.  In
fact, we saw there that the enhanced tau
production directly led to a substantial contribution to OSOF.  Still,
there is hope that the prompt OSOF
dileptons will have a distinct enough distribution to ultimately stand out
against the low-mass tau shape.
Even if this is not so clean, the complete set of signals would still be
quite consistent with a tau-sneutrino NLSP.

In a broader context, it is clear that we have found some very distinctive
new phenomenology within the MSSM.  These signals 
have been largely overlooked in the past, partly because sneutrino NLSP spectra 
do not readily appear in canonical UV mediation scenarios, and partly because
there has been little pressure to model-build from the dark matter perspective.
However, there is obviously a benefit in exploring a more flexible approach to the MSSM, 
especially given our
near-ignorance of physics beyond even a few hundred GeV.  Conversely, should our signals be
discovered in the coming years, they would indicate a SUSY mediation scenario quite distinct 
from what is typically imagined.

\acknowledgments{ We are grateful to
Kaustubh Agashe, Maxim Perelstein, Yael Shadmi and especially Matt Reece for discussions 
and help with the {\tt BRIDGE} program. We are also grateful to Matt Reece and Yael Shadmi 
for useful comments on the manuscript.
A.K. is partially supported by NSF under grant PHY-0801323.  B.T. is supported by the Leon
Madansky Fellowship and Johns Hopkins University grant \#80020033.

\appendix

\section{Formulas for RH Slepton 3-Body Decay Widths}\label{decayrate}

Here we present the 3-body decay widths for the RH sleptons via an intermediate off-shell Bino.
These can become the dominant modes when the charginos and neutralinos are all heavier, so that
2-body modes $\sel^+ \to e_R^+ \; \gaugino^0$ (e.g., into mostly-Wino via gaugino mixing) are
cut off.  They will compete with the modes induced by left-right mixing, as discussed
in subsection~\ref{lrmixing}.

There are two types of 3-body processes:  $\sel^+ \to e_R^+ \, l_L^+ \,\slep_L^-$ proceeding
via a chirality-preserving $\bino$ propagator, and $\sel^+ \to e_R^+ \, l_L^- \, \slep_L^+$
proceeding via a chirality-flipping $\bino$ propagator.  The general expressions for the widths
of these decays depend on the three parameters $\me$, $\ml$, and $\mB$, and are somewhat awkward.
We also include the limits where $(\me-\ml) \to 0$ or $\ml \to 0$, namely where the slepton is
respectively nonrelativistic (NR) or ultrarelativistic (UR) in the $\sel$ rest frame.  We also
display the $\mB \gg \me$ limit (non-propagating Bino) for the decays, since the expressions
significantly simplify.

There are also the equivalent processes with $l$ replaced by $\nu$:  $\sel^+ \to e_R^+\, \bar \nu\, \tilde \nu$
proceeding via a chirality-preserving $\bino$ propagator, and $\sel^+ \to e_R^+\; \nu\; \tilde \nu^*$
proceeding via a chirality-flipping $\bino$ propagator.  The expressions are identical, up
to the obvious replacement $m_{\slep}\to m_{\snu}$.

We ignore gaugino mixing effects, treating the Bino as a pure mass state.  Extending the formulas
to a mixed neutralino is straightforward.

$g_1$ stands for the $U(1)_Y$ coupling ($g_1 \simeq 0.37$), and $Y_i$ stands
for the hypercharge ($\Ye = 1$ and $\Yl = -1/2$).

\footnotesize{
\beqarray
\lefteqn{\!\!\!\!\!\!\!\!\!\!\!\!\!\!\!\! \mbox{\normalsize $\Gamma(\sel^+ \to e_R^+ \; l_L^+ \; \slep_L^-)\;$ (chirality-preserving): }} \\
\nonumber \\
{\rm Full} &   = & \frac{\Ye^2\Yl^2g_1^4}{2^8\pi^3} \frac{1}{\mB^4\me^3} \left[  \mB^2(\me^2-\ml^2)\left(6\mB^4+2\me^2\ml^2-5\mB^2(\me^2+\ml^2)\right)  +  2\me^4\ml^4\log\left(\frac{\me^2}{\ml^2}\right) +  \right. \nonumber \\
           &     & \left. 2(\mB^2-\ml^2)\left(3\mB^6 + \mB^2\me^4 + \me^4\ml^2 - \mB^4(4\me^2+\ml^2)\right)\log\left(\frac{\mB^2-\me^2}{\mB^2-\ml^2}\right) \right] \nonumber \\
{\rm NR}   &   = & \frac{\Ye^2\Yl^2g_1^4}{ 3\cdot 2^5\pi^3} \frac{\me}{\ml} \left[ 2\frac{\me-\ml}{\me^4}\left(3(\mB^2-\me^2)^2+6(\mB^2-\me^2)\me(\me-\ml)+\me^2(\me-\ml)^2\right)  -   \vphantom{\left(\frac{\me^2}{\ml^2}\right)}  \right.  \nonumber \\
           &     & 3\frac{\mB^2-\me^2}{\me^5}\left((\mB^2-\me^2)^2+3(\mB^2-\me^2)\me(\me-\ml)+2\me^2(\me-\ml)^2\right)\times \nonumber \\
           &     & \left. \log\left(1+\frac{2\me(\me-\ml)}{\mB^2-\me^2}\right)  \right] \nonumber \\ 
           & \to & \frac{\Ye^2\Yl^2g_1^4}{15\cdot 2^3\pi^3} \frac{(\me-\ml)^5}{(\mB^2-\me^2)^2} \frac{\me}{\ml} \nonumber \\ 
{\rm UR}   &   = & \frac{\Ye^2\Yl^2g_1^4}{2^8\pi^3} \me \left[ 6\frac{\mB^2}{\me^2} -5 + 2\left( 3\frac{\mB^4}{\me^4} - 4\frac{\mB^2}{\me^2} + 1 \right) \log\left( 1 - \frac{\me^2}{\mB^2} \right) \right] \nonumber \\
           & \to & \frac{\Ye^2\Yl^2g_1^4}{3\cdot2^9\pi^3} \frac{\me^5}{\mB^4} \nonumber \\
\nonumber \\
\nonumber \\
\lefteqn{\!\!\!\!\!\!\!\!\!\!\!\!\!\!\!\! \mbox{\normalsize $\Gamma(\sel^+ \to e_R^+ \; l_L^- \; \slep_L^+)\;$ (chirality-flipping): }} \\
\nonumber \\
{\rm Full} &   = & \frac{\Ye^2\Yl^2g_1^4}{2^7\pi^3} \frac{1}{\mB^4\me^3} \left[  \mB^2(\me^2-\ml^2)\left(2\mB^4 + 2\me^2\ml^2 - \mB^2(\me^2+\ml^2)\right) + \vphantom{\left(\frac{\me^2}{\ml^2}\right)} \right.  \nonumber \\
           &     &  2\me^2\ml^2\left(\me^2\ml^2 - \mB^2(\me^2+\ml^2)\right)\log\left(\frac{\me^2}{\ml^2}\right) + \nonumber \\
           &     &  \left. 2(\mB^2-\me^2)(\mB^2-\ml^2)(\mB^4-\me^2\ml^2)\log\left(\frac{\mB^2-\me^2}{\mB^2-\ml^2}\right)  \right]  \nonumber \\
{\rm NR}   &   = & \frac{\mB^2}{\me^2} \, \Gamma(\sel^+ \to e_R^+ \; l_L^+ \; \slep_L^-)_{\rm NR} \nonumber \\
{\rm UR}   &   = & \frac{\Ye^2\Yl^2g_1^4}{2^7\pi^3} \frac{\mB^2}{\me} \left[ 2 - \frac{\me^2}{\mB^2} + 2\left(\frac{\mB^2}{\me^2} - 1 \right) \log\left( 1 - \frac{\me^2}{\mB^2} \right) \right]  \nonumber \\
           & \to & \frac{\Ye^2\Yl^2g_1^4}{3\cdot2^7\pi^3} \frac{\me^3}{\mB^2} \nonumber
\eeqarray
}

\bibliography{lit}
\bibliographystyle{apsper}
\end{document}